\documentclass[aps,groupedaddress]{revtex4}

\usepackage{graphicx}

\newcommand{\rr}{{\bf r}}

\begin{document}

\title{Perfect Optical Solitons: Spatial Kerr Solitons as Exact Solutions of Maxwell's Equations}

\author{Alessandro Ciattoni}
\address{Istituto Nazionale per la Fisica della Materia, UdR Universit\'a dell'Aquila, 67010, L'Aquila, Italy, and
\\ Dipartimento di Fisica, Universit\'a dell'Aquila, 67010 L'Aquila, Italy}

\author{Bruno Crosignani}
\address{Dipartimento di Fisica, Universit\'a dell'Aquila, 67010 L'Aquila, Italy and \\ Istituto Nazionale di
Fisica della Materia, UdR Roma "La Sapienza", 00185 Roma, Italy and \\ California Institute of Technology 128-95,
Pasadena, California 91125}

\author{Paolo Di Porto}
\address{Dipartimento di Fisica, Universit\'a dell'Aquila, 67010, L'Aquila, Italy and \\ Istituto Nazionale di
Fisica della Materia, UdR Roma "La Sapienza", 00185 Roma, Italy}

\author{Amnon Yariv}
\address{California Institute of Technology 128-95, Pasadena, California 91125}

\begin{abstract}
We prove that spatial Kerr solitons, usually obtained in the frame of nonlinear Schr\"{o}dinger  equation valid in
the paraxial approximation, can be found in a generalized form as exact solutions of Maxwell's equations. In
particular, they are shown to exist, both in the bright and dark version, as linearly polarized exactly integrable
one-dimensional solitons, and to reduce to the standard paraxial form in the limit of small intensities. In the
two-dimensional case, they are shown to exist as azimuthally polarized circularly symmetric dark solitons. Both
one and two-dimensional dark solitons exhibit a characteristic signature in that their asymptotic intensity cannot
exceed a threshold value in correspondence of which their width reaches a minimum subwavelength value.
\end{abstract}

\maketitle

\section{Introduction}
The analytic description of spatial Kerr solitons, initiated by the seminal paper of Chiao et al.,\cite{Chiao} has
been continuously evolving in the last forty years.\cite{Trillo,Kivshar} It basically hinges upon the use of the
nonlinear Schr\"{o}dinger equation (NLS), which in turn follows from the nonlinear Helmholtz equation once the
paraxial approximation, limiting the size $\sigma$ of the propagating beam to values large compared to the
wavelength $\lambda$, is introduced. This approximation becomes inappropriate if the beam size $\sigma$ is
comparable with $\lambda$, a regime where nonparaxial effects become important and are eventually able to provide
a mechanism for avoiding nonphysical behaviors (like,e.g., catastrophic collapse) in the beam evolution. Although
many contributions have been produced in this direction,\cite{sca1,sca2,sca3,sca4,vec1,vec2,vec3,coq1,coq2,coq3}
they are typically based on some form of asymptotic expansion in the smallness parameter $\eta = \lambda / \sigma$
and are thus limited to the range $\eta < 1$. To overcome this limitation, we start {\it ab initio} from Maxwell's
equations and look for exact soliton solutions. More precisely, we solve Maxwell's equations in the presence of a
fully vectorial Kerr polarizability and find a class of {\it perfect optical solitons} which inherently include
all nonparaxial contributions. This is separately performed for one dimensional and two dimensional spatial
solitons, both in the bright and dark configuration. In particular, the one dimensional case is dealt with by
reducing Maxwell's equations to a system of first order differential equations and handling it by appealing to the
usual formalism employed in the frame of dynamical systems. Our system is shown to posses a first integral so that
its integrability is proved and the boundary value problem, associated with solitons, solved in closed analytical
form.

One the main results obtained in this paper is the proof of the existence of exact solutions of Maxwell equations
in the form of linearly polarized {\it one dimensional} Kerr solitons: they do not suffer of any limitation on the
value of $\sigma$ and $\lambda$ (apart from the obvious ones associated with the validity of the macroscopic model
of Kerr polarizability) and their existence curve can be numerically evaluated for all values of the beam
intensity. Both bright and dark solitons can be derived from an integrable system of equations and their existence
curve shows that, in the case of bright solitons, any value of the peak intensity $u_{x0}^2$ is allowed, while
dark solitons can only exist if their asymptotic intensity $u_{x\infty}^2$ does not exceed a threshold value
completely determined by Kerr coefficients. In correspondence to this threshold, their width approaches the
minimum value of the order of a fraction of $\lambda$. In the {\it two dimensional} case, dark azimuthally
polarized solitons are found and their existence curve implies, the same threshold behavior of the one dimensional
dark solitons. While one dimensional solitons reduce to the standard paraxial ones for small values of the
intensity, the two dimensional azimuthal dark soliton is a completely new entity which has never been studied in
the paraxial regime.

We wish to note that the proof of the existence and derivation of exact solitons requires, in the one dimensional
case, the use of a rather sophisticated mathematical analysis borrowed from the dynamical system formalism, which
we decide to report in full in section 2.
\section{One dimensional spatial solitons}
The electric and magnetic complex amplitudes ${\bf E}(\rr)$ and ${\bf B}(\rr)$ of a monochromatic electromagnetic
field $Re[{\bf E} \exp(-i \omega t)]$, $Re[{\bf B} \exp(-i \omega t)]$ propagating in a nonlinear medium obey
Maxwell's equations
\begin{eqnarray} \label{maxwell1}
\nabla \times {\bf E} &=& i \omega {\bf B}, \nonumber \\
\nabla \times {\bf B} &=& -i \frac{\omega}{c^2} n_0^2 {\bf E} - i \omega \mu_0 {\bf P}_{nl},
\end{eqnarray}
where $n_0$ labels the linear refractive index and ${\bf P}_{nl}$ is the nonlinear polarizability. In the case of
nonresonant isotropic media \cite{Cros}, the vectorial Kerr effect is described by the polarizability
\begin{equation} \label{polariz}
{\bf P}_{nl} = \frac{4}{3} \epsilon_0 n_0 n_2 \left[ |{\bf E}|^2 {\bf E} + \frac{1}{2} ({\bf E \cdot E}) {\bf E}^*
\right],
\end{equation}
$n_2$ being the nonlinear refractive index coefficient. After eliminating $\bf B$ from Eq.(\ref{maxwell1}) and
takind advantage of Eq.(\ref{polariz}), we get
\begin{equation} \label{maxwell2}
\nabla \times \nabla \times {\bf E} =  k^2 {\bf E} + k^2 \frac{4}{3} \frac{n_2}{n_0}  \left[ |{\bf E}|^2 {\bf E} +
\frac{1}{2} ({\bf E \cdot E}) {\bf E}^* \right]
\end{equation}
where $k = n_0 \omega/c$. We now introduce a Cartesian reference frame $Oxyz$ with unit vectors $\hat{\bf
e}_x$,$\hat{\bf e}_y$,$\hat{\bf e}_z$, and look for one dimensional solitons propagating along the $z-$ axis, that
is for $y-$ independent fields of the form
\begin{equation} \label{soliton}
{\bf E}(x,y,z) = \exp (i \alpha z) \left[U_x (x) \hat{\bf e}_x + i U_z (x) \hat{\bf e}_z \right]
\end{equation}
where $U_x$ and $U_z$ depends on $x$ alone and $\alpha$ is a real constant. Substituting Eq.(\ref{soliton}) into
Eq.(\ref{maxwell2}) yields the system of ordinary differential equations
\begin{eqnarray} \label{system1}
\alpha \frac{d U_z}{dx} = \left[(\alpha^2-k^2) - \frac{2 k^2 n_2}{n_0} \left( U_x^2 + \frac{1}{3} U_z^2 \right)
\right] U_x, \nonumber \\
\frac{d^2 U_z}{dx^2} - \alpha \frac{d U_x}{dx} = - k^2 \left[ 1 + \frac{2 n_2}{n_0}  \left( \frac{1}{3} U_x^2 +
U_z^2 \right)\right] U_z,
\end{eqnarray}
whose unknowns $U_x$ and $U_z$ are real (as a consequence of the $\pi / 2$ phase difference we introduced between
the transverse and longitudinal field components (see Eq.(\ref{soliton})). Note that the field in
Eq.(\ref{soliton}) has a vanishing $y-$component, a requirement not forbidden by Maxwell's equations. From
Eqs.(\ref{system1}), it is also evident that the $z-$ component $U_z$ only vanishes if $U_x = \pm \sqrt{(n_0
/2n_2)(\alpha^2/k^2 - 1)}$ which describes a family of solitary plane waves rather than solitons \cite{Chen}. The
fact that $U_z$ does not generally vanish is a consequence of the vectorial coupling between transverse and
longitudinal components which cannot be rigorously neglected when describing spatially nonuniform fields, like for
examples solitons (From a physical point of view, this follows from the first Maxwell equation setting the
divergence of the electric field). Note that the longitudinal component is usually neglected in the paraxial
regime thanks to the slow variation of the transverse component as compared to the wavelength $\lambda = 2 \pi /
k$, a circumstance which allows to treat it as a perturbation for slightly nonparaxial beams \cite{NPB1,NPB2}. In
the present paper, we deal on equal foot with both transverse and longitudinal components and it is their
simultaneous non-vanishing and coupling which allows us to find exact solitons.

Equations (\ref{system1}) can be recast in a more symmetric form by differentiating the first one and consequently
eliminating $d^2 U_z / dx^2$ (together with $d U_z / dx$) from the second one, thus getting
\begin{eqnarray} \label{system2}
\frac{d u_x}{d\xi} = \frac{\left[ \beta^2 \left( 1 - \frac{2}{3} \gamma u_x^2 +2 \gamma u_z^2 \right)+
\frac{4}{3}\left(\gamma + 2 u_x^2 + \frac{2}{3}u_z^2 \right) u_x^2 \right]}
{\beta \left[ 1 + \gamma \left(6 u_x^2 + \frac{2}{3} u_z^2 \right) \right]}
u_z \equiv Q_x (u_x, u_z |\beta ), \nonumber \\
\frac{d u_z}{d \xi} = \frac{1}{\beta} \left[(\beta^2-1) - 2 \gamma \left( u_x^2 + \frac{1}{3} u_z^2 \right)
\right] u_x \equiv Q_z(u_x,u_z |\beta),
\end{eqnarray}
where we have introduced the dimensionless variables $\xi = k x$, $\beta = \alpha /k$ and $(u_{x} , u_{z}) =
\sqrt{|n_2|/n_0} (U_{x} , U_{z})$, while $\gamma = n_2 / |n_2|$ (so that $\gamma = + 1$ and $\gamma = -1$ for
focusing and defocusing media, respectively). Equations (\ref{system2}) are a system of first order differential
equations describing any electromagnetic field of the form of Eq.(\ref{soliton}) and they are equivalent to
Maxwell's equation, providing the relation
\begin{equation} \label{noncrossellipse}
1 + \gamma \left( 6 u_x^2 + \frac{2}{3} u_z^2 \right) \neq 0
\end{equation}
uniformly (i.e. for any $\xi$) holds. Equations (\ref{system2}) can be conveniently regarded as an autonomous
dynamical system (since $Q_x$ and $Q_z$ does not explicitly depend upon $\xi$), whose solutions, or orbits, are
$\xi-$parameterized curves ${\bf u} (\xi) = (u_x(\xi) \: u_z(\xi))^T$ (belonging to the phase plane $(u_x,u_z)$),
tangent at each point to the vector field ${\bf Q} = (Q_x \: Q_z)^T$. Solitons are particular orbits which, for
suitable values of $\beta$, pass through two special points of the phase plane imposed by the boundary conditions
pertinent to each soliton kind (boundary value problem).

The most remarkable and general property of the system of Eqs.(\ref{system2}) is that it is a {\it conservative}
system, i.e., it admits a {\it first integral} $F(u_x,u_z|\beta)$, defined over the phase plane, satisfying the
relation
\begin{equation} \label{FirstInt}
0 = \frac{dF}{d\xi} \equiv \frac{\partial F}{\partial u_x} \frac{d u_x}{d \xi }+\frac{\partial F}{\partial u_z}
\frac{d u_z}{d \xi } =  \frac{\partial F}{\partial u_x} Q_x + \frac{\partial F}{\partial u_z} Q_z.
\end{equation}
In fact, it is straightforward to prove that the function
\begin{equation} \label{FF}
F(u_x,u_z | \beta) = 2 u_x^6 + \frac{4}{3} u_x^4 u_z^2 + \frac{2}{9} u_x^2 u_z^4 - \frac{1}{2} \gamma (3 \beta^2
-4) u_x^4 + \frac{1}{3} \gamma (2-\beta^2) u_x^2 u_z^2 + \frac{1}{2} \gamma \beta^2 u_z^4 - \frac{1}{2}(\beta^2-1)
u_x^2 + \frac{1}{2} \beta^2 u_z^2
\end{equation}
obeys Eq.(\ref{FirstInt}) whenever Eq.(\ref{noncrossellipse}) is satisfied. This implies that $F$ is a first
integral of the system of Eqs.(\ref{system2}) whenever this system is equivalent to Maxwell's equations. According
to Eq.(\ref{FirstInt}), any solution of Eqs.(\ref{system2}) is constrained to move along a single level set
\begin{equation} \label{levelset}
F(u_x,u_z | \beta)=F_0.
\end{equation}
Inverting Eq.(\ref{levelset}) furnishes $u_z = u_z(u_x,F_0,\beta)$ which, once inserted into the first of
Eqs.(\ref{system2}), yields a first order differential equation solvable by quadratures, thus proving the {\it
integrability} of Eqs.(\ref{system2}). Note that the first integral in Eq.(\ref{FF}) contains even powers of $u_x$
and $u_z$ only so that any level set of Eq.(\ref{levelset}) is invariant under the inversion of the phase plane
$(u_x,u_z) \rightarrow -(u_x,u_z)$.

Exploiting the properties of the first integral found above, we are now in a position to solve in a direct way the
soliton boundary value problem, that is to find suitable values of $\beta$ (if any) for which a solution
$u_x(\xi),u_z(\xi$) of Eqs.(\ref{system2}) satisfies the general boundary conditions
\begin{equation} \label{bounCon}
\left( \begin{array}{c} u_x(0) \\ u_z(0) \end{array} \right)  = \left( \begin{array}{c} u_{x0} \\
u_{z0}\end{array} \right) \equiv {\bf u}_0,
\left( \begin{array}{c} u_x(+\infty) \\ u_z( +\infty) \end{array} \right)  = \left( \begin{array}{c} u_{x\infty} \\
u_{z \infty}\end{array} \right) \equiv {\bf u}_\infty.
\end{equation}
where ${\bf u}_0$ and ${\bf u}_{\infty}$ are defined by the kind of soliton, bright or dark, we wish to consider.
From a geometrical point of view, this implies that the associated integral curve on the phase plane $(u_x,u_z)$
has to pass through the points ${\bf u}_0$ and ${\bf u}_{\infty}$, or, using Eq.(\ref{levelset}),
\begin{eqnarray} \label{Fnec}
F(u_{x0} , u_{z0} | \beta) = F_0 , \nonumber \\
F(u_{x \infty} , u_{z \infty} | \beta) = F_0.
\end{eqnarray}
Since ${\bf u}_\infty$ has to be reached for $\xi \rightarrow +\infty$, it is obvious that ${\bf u}_{\infty}$ has
to be an equilibrium point of Eqs.(\ref{system2}), that is
\begin{eqnarray} \label{Qnec}
Q_x(u_{x \infty} , u_{z \infty} | \beta) = 0 , \nonumber \\
Q_z(u_{x \infty} , u_{z \infty} | \beta) = 0.
\end{eqnarray}
Equations (\ref{Fnec}) and (\ref{Qnec}) in the unknowns $\beta$, $F_0$ are necessary conditions for the solitons
existence. They become also sufficient if, once $\beta$ and $F_0$ are determined, one is able to prove that the
integral curve actually reaches the point ${\bf u}_\infty$. Following the outlined procedure, the existence of
both bright and dark solitons will be proved and the corresponding existence conditions and propagation constants
$\beta$ will be found.
\subsection{Bright Solitons}
Bright solitons are localized nondiffracting beams, that is solutions of Eqs.(\ref{system2}) vanishing for $|\xi|
\rightarrow +\infty$, which in turn requires ${\bf u}_\infty = 0$. Note that Eqs.(\ref{Qnec}) are automatically
satisfied by this boundary condition since the origin $(u_x,u_z) = (0,0)$ is always an equilibrium point of
Eqs.(\ref{system2}). The second of Eqs.(\ref{Fnec}) directly gives $F_0 = 0$ so that the remaining condition we
have to impose is the first of Eq.(\ref{Fnec}) that is
\begin{equation} \label{BriCon}
F(u_{x0} , u_{z0} | \beta) = 0.
\end{equation}
In order to set the boundary condition ${\bf u}_0$ we note that, because of the invariance of the level set in
Eq.(\ref{levelset}) under inversion of the phase plane, a soliton has to be associated with an integral curve
starting from and ending into the origin and that this curve has to be symmetric under either the reflection $u_x
\rightarrow -u_x$ or the reflection $u_z \rightarrow -u_z$. Because of these symmetry properties, we have ${\bf
u}_0 = (0 \: u_{z0})^T$ and ${\bf u}_0 = (u_{x0} \: 0)^T$ (where the symbol $T$ stands for transposed) in the
former and in the latter case, respectively. In the first case, Eq.(\ref{BriCon}) becomes $\beta^2 u_{z0}^2
(\gamma u_{z0}^2 +1) = 0$ which implies $\beta = 0$, so that soliton propagation is not allowed. We are left to
consider the case ${\bf u}_0 = (u_{x0} \: 0)^T$ for which Eq.(\ref{BriCon}) furnishes
\begin{equation} \label{betasq}
\beta ^2 = \frac{(1 + 2\gamma u_{x0}^2)^2}{1+3 \gamma u_{0x}^2}.
\end{equation}
In appendix A, we prove that bright solitons exist for all the real values of $u_{x0}$ in focusing media ($\gamma
= 1$) and that they never exist in defocusing media ($\gamma = -1$) in agreement with the intuitive behavior of
Kerr nonlinearity which tends to tighten and to spread the beam in these two cases, respectively. Obviously, the
above results about the existence of bright solitons are based on the validity of Eq.(\ref{polariz}), which fails
either for large intensities or for soliton widths so small to invalidate the continuum description of the
material response. For $\gamma = 1$, Eq. (\ref{betasq}) yields
\begin{equation} \label{betabright}
\beta  = \pm \frac{1 + 2 u_{x0}^2}{\sqrt{1+3 u_{0x}^2}}
\end{equation}
which is the propagation constant of the exact bright solitons. The double sign in Eq.(\ref{betabright}) describes
the two counter-propagating solitons along the $z-$axis.

Substituting Eq.(\ref{betabright}) and $F_0=0$ into Eq.(\ref{levelset}), we obtain the equation for the integral
curves on the phase plane corresponding to bright solitons, and these are reported, for some values of $u_{x0}$,
in Figure 1. Note that, for each $|u_{x0}|$, the corresponding level set is a bow-tie shaped curve encompassing
three orbits of Eqs.(\ref{system2}), that is the origin (which is an equilibrium point) and the left and right
loop of the bow-tie. These last two orbits correspond to a pair of bright solitons each of which can be obtained
from the other after the inversion of the $x-$axis, $\xi \rightarrow -\xi$ (implying the reflection $u_x
\rightarrow -u_x$ also), as expected because of the reflection invariance along any directions shown by Kerr
nonlinearity. Considering the right half plane $u_x > 0$ only, we observe that soliton curve $u_x(\xi),u_z(\xi)$
explore the loop starting form the origin (for $\xi=-\infty$), reaching the point $(u_x,u_z)=(u_{x0},0)$ (for
$\xi=0$) and ending into the origin (for $\xi=+\infty$). From Eqs.(\ref{system2}) it is evident that the loop is
explored counter-clockwise and clockwise for $\beta > 0$ and $\beta < 0$, respectively, so that, for
counter-propagating solitons (denoted with $(+)$ and $(-)$), we have $u_x^{(+)}(\xi) = u_x^{(-)}(\xi)$ and
$u_z^{(+)}(\xi) = -u_z^{(-)}(\xi)$.

Having proved the bright soliton existence and derived the associated propagation constant $\beta$, we are now in
the position to obtain the soliton shape for any given $u_{x0}$ by numerically solving Eqs.(\ref{system2}) with
$\beta$ given by Eq.(\ref{betabright}) and the initial conditions $u_x(0)=u_{x0}$, $u_z(0)=0$ (the numerical
approach being much simpler than integrating the system Eqs.(\ref{system2}) by quadrature). In Figure 2, we report
the plots of the transverse $u_x$ and longitudinal $u_z$ components of the bright solitons for the same $u_{x0}$
as in Figure 1. Note that, as expected, the soliton width decreases for increasing $u_{x0}$, while the
longitudinal component $u_z$ increases. In Figure 3, we report the bright soliton existence curve, relating the
FWHM ($\Delta_{bright}$) to $|u_{x0}|$. As $|u_{x0}|$ decreases the width indefinitely increases and diverges for
$|u_{x0}|=0$; on the contrary, as $|u_{x0}|$ increases, the width decreases monothonically approaching zero.
\subsection{Dark Solitons}
In the scalar approximation, dark solitons are nondiffrating beams vanishing at $\xi = 0$ and approaching an
asymptotic amplitude value for $|\xi| \rightarrow +\infty$. In our vectorial case, the natural extension of the
previous definition is identified with soliton solutions with ${\bf u}_0 = (0 \: \: u_{z0})^T$, and ${\bf
u}_\infty = (u_{z\infty} \: \: 0)^T$ (see Eqs.(\ref{bounCon})). In fact, the above boundary conditions will be
proved to describe an exact dark soliton which, in the paraxial limit, reduces to the standard scalar dark one.

The chosen values of ${\bf u}_\infty$ identically satisfy the first of Eq.(\ref{Qnec}). The second of
Eqs.(\ref{Qnec}) implies, with the help of the second of Eqs.(\ref{system2}),
\begin{equation} \label{q1}
\beta^2 = 1 + 2 \gamma u_{x \infty}^2.
\end{equation}
Substituting this value of $\beta^2$ together with the boundary conditions into Eqs.(\ref{Fnec}), we get
\begin{eqnarray} \label{q2}
F_0 = -\frac{\gamma}{2}(1+2\gamma u_{x\infty}^2) u_{x\infty}^4, \nonumber \\
u_{x \infty}^4 = -\gamma u_{z0}^2-u_{z0}^4.
\end{eqnarray}
The first of these equations furnishes the value $F_0$ of the first integral along the dark soliton integral
curve. The second one is a necessary condition for soliton existence from which we immediately obtain $\gamma =
-1$, in agreement with the intuitive property that only defocusing media can support dark solitons. In Appendix B,
we prove that dark solitons exist for $u_{x\infty}^2 < 1/6$ only and that
\begin{eqnarray} \label{darkexist}
\beta = \pm \sqrt{1 - 2 u_{x \infty}^2}, \nonumber \\
u_{z0} = \pm \sqrt{\frac{1}{2}\left( 1 - \sqrt{1-4 u_{x\infty}^2} \right)}, \nonumber \\
F_0 = \frac{1}{2} \left(1 - 2 u_{x \infty}^2\right) u_{x\infty}^4,
\end{eqnarray}
so that each soliton is completely specified by the value $u_{x\infty}$ only.

As in the case of bright solitons, the integral curves in the phase plane associated to dark solitons are given by
Eq.(\ref{levelset}), with $\beta$ and $F_0$ given in Eqs.(\ref{darkexist}), some of them being reported in Figure
4. For each $|u_{x\infty}|$ the level set is a closed curve encompassing four orbits of Eqs.(\ref{system2}) that
is the two equilibrium points $(-u_{x\infty},0)$ and $(u_{x\infty},0)$ together with the two curves joining these
two points in the upper and lower half plane, respectively. These last two orbits are associated to a pair of dark
solitons having opposite longitudinal components. Limiting our attention to the upper half plane $u_z > 0$, the
dark soliton curve $u_x(\xi),u_z(\xi)$ starts, for $\beta > 0$, from the point $(-u_{x\infty},0)$ at $\xi =
-\infty$, reaches the point $(0,u_{z0})$ at $\xi=0$ and finally ends into the point $(u_{x\infty},0)$ at $\xi =
+\infty$ (for $\beta < 0$ it is sufficient to invert $\xi \rightarrow -\xi$).

For any given value of $u_{x\infty}$ (in the range $|u_{x\infty}|<1/\sqrt{6}$), the shape of dark solitons can be
obtained by numerically integrating Eqs.(\ref{system2}) with $\beta$ given by the first of Eqs.(\ref{darkexist})
and initial conditions $u_x(0)=0$ and $u_z(0)=u_{z0}$ (the latter being given by the second of
Eqs.(\ref{darkexist})). In Figure 5, we plot the transverse $u_x$ and longitudinal $u_z$ components of various
dark solitons, for the same $u_{x\infty}$ as in Figure 4. Also in this case, for increasing $u_{x\infty}$ the
soliton width decreases while the longitudinal component increases. In Figure 6 we report the dark soliton
existence curve relating the soliton FWHM ($\Delta_{dark}$) to $u_{x\infty}$, in the range
$0<u_{x\infty}<1/\sqrt{6}$. Note that, for very small $u_{x\infty}$. the FWHM indefinitely grows whereas in
correspondence to the threshold value $u_{x\infty}=1/\sqrt{6}$, it attains its minimum value $\simeq 4$,
corresponding to dimensional value $\simeq (2/\pi) \lambda \simeq 0.63 \lambda$.
\subsection{The Optical Intensity}
Having derived the electric field (see Eq.(\ref{soliton})) associated to both bright and dark solitons, we can
directly evaluate the corresponding magnetic field by means of the pertinent Maxwell equation. Substituting
Eq.(\ref{soliton}) into the first of Eqs.(\ref{maxwell1}) we easily deduce, in terms of the dimensionless fields,
\begin{equation} \label{Magnetic}
{\bf B}(x,y,z) = \frac{k}{\omega} \sqrt{\frac{n_0}{|n_2|}} \exp(i\beta kz) \left( \beta u_x - \frac{d u_z}{d \xi}
                 \right)_{\xi=kx} \hat{\bf e}_y.
\end{equation}
Note that the soliton magnetic field is parallel to the $y-$axis and therefore orthogonal to the electric field
everywhere, a remarkable vectorial feature that exact solitons shares with plane waves (which are rigorously
nondiffracting fields as well). In order to describe the soliton energy flow, we can now evaluate the averaged
Poynting vector ${\bf S} = Re ({\bf E \times B^*}) / (2 \mu_0)$ which, using Eqs.(\ref{soliton}) and
(\ref{Magnetic}) and the second of Eqs.(\ref{system2}), becomes
\begin{equation} \label{avpoy}
{\bf S} = \frac{I_0}{\beta} \left[1 + 2\gamma \left(u_x^2 + \frac{1}{3} u_z^2 \right) \right] u_x^2 \hat{\bf e}_z
          \equiv \frac{\beta}{|\beta|} I \hat{\bf e}_z,
\end{equation}
where $I_0 = k n_0 /(2\omega \mu_0 |n_2|)$ and $I$, the modulus of the averaged Poynting vector, is the optical
intensity. The averaged Poynting vector lies along the $z-$axis everywhere and this is fully consistent with the
nondiffracting nature of the solitons we are considering (which is not rigorously the case in the paraxial
approximation). Note that $\bf S$ is proportional to $\beta ^ {-1}$ and the expression in square brackets of
Eq.(\ref{avpoy}) is always positive (while this is trivial in the case $\gamma =+1$, in the case $\gamma = -1$ all
the orbits $u_x(\xi),u_z(\xi)$ of Eqs.(\ref{system2}) lie inside the ellipse defined in
Eqs.(\ref{noncrossellipse}), which is in turn contained within the ellipse $2\left(u_x^2 + \frac{1}{3} u_z^2
\right)=1$, so that the expression in square bracket of Eq.(\ref{avpoy}) is always positive). This implies, as
expected, that, both for bright and dark solitons, the sign of $\beta$ determines whether ${\bf S}$ is parallel or
antiparallel to the $z-$axis. In Figure 7 we report the plots of the normalized optical intensity $I/I_0$ for the
same bright and dark solitons examined in the previous Figures. From Eq.(\ref{avpoy}) we observe that the optical
intensity is in general not proportional to the square modulus of the electric field. However, in the paraxial
limit where $u_x << 1$, $u_z << u_x$ and $\beta^{-1} \simeq 1$, Eq.(\ref{avpoy}) gives $I = I_0 u_x^2$,
reproducing the well-known result typical of paraxial optics. We can also evaluate the maximum soliton optical
intensity, that is Eq.(\ref{avpoy}) at $\xi = 0$ (and $\gamma = +1$) for bright solitons and at $|\xi| = +\infty$
(and $\gamma = -1$) for dark solitons, thus getting
\begin{eqnarray} \label{maxint}
I_{bright} = I_0 \sqrt{1+3 u_{x0}^2} u_{x0}^2, \nonumber \\
I_{dark} = I_0 \sqrt{1 - 2u_{x\infty}^2} u_{x_\infty}^2.
\end{eqnarray}
From these equations we note that $I_{bright} > I_0 u_{x0}^2$ whereas $I_{dark} < I_0 u_{x\infty}^2$ so that, in
general, bright and dark solitons are characterized by an optical intensity which is greater and smaller,
respectively, than the corresponding paraxial prediction. This is evidently associated with the fact that, in an
extremely narrow soliton, the longitudinal component of the electric field is as large as the transverse one.
\subsection{Paraxial limit}
The above description of one dimensional bright and dark solitons is exact, no approximation having been exploited
in their analytical derivation. As a consequence, the solitons described above are expected to reduce, in the
paraxial limit where the soliton width is much larger than the wavelength, to those predicted by the NLS. The
paraxial limit is clearly obtained by considering the range of values
\begin{eqnarray} \label{parlim}
u_x << 1, \nonumber \\
u_z << u_x,
\end{eqnarray}
since the soliton width increases for decreasing optical intensities while the longitudinal component decreases.
By differentiating the first of Eqs.(\ref{system2}), using the second of Eqs.(\ref{system2}) to eliminate $du_z /d
\xi$ and exploiting Eqs.(\ref{parlim}) to retain only the first relevant order, we obtain
\begin{equation} \label{parpar}
\frac{d^2 u_x}{d \xi^2} = (\beta^2 -1) u_x - 2 \gamma u_x^3.
\end{equation}
Note that, in describing paraxial Kerr solitons, the electric field is usually expressed as $E_x (x,z) = \exp [i k
( 1+\widetilde{\beta}) z] \sqrt{n_0/|n_2|} u_x (\xi)$, where the fundamental plane wave carrier $\exp(i k z)$ is
separated by the slowly varying amplitude of the field. The comparison of this field expression with
Eq.(\ref{soliton}) yields $\beta = 1 + \widetilde{\beta}$ with $\widetilde{\beta} << 1$ (consisting with the
paraxial picture where the main plane wave carrier is slowly modulated), implying that $\beta^2 - 1 \simeq 2
\widetilde{\beta}$. Introducing this relation into Eq.(\ref{parpar}), we get
\begin{equation} \label{nlssol}
-\widetilde{\beta} u_x+\frac{1}{2} \frac{d^2 u_x}{d \xi^2} = - \gamma u_x^3,
\end{equation}
which coincides with the usual equation (obtained from the NLS) describing paraxial Kerr solitons.  Equation
(\ref{nlssol}) admits both of bright soliton solutions ($\gamma = +1$) of the form $u_x (\xi) = u_{x0} \:
\textrm{sech} ( u_{x0} \xi )$ and of dark soliton solutions ($\gamma = -1$) of the form $u_x (\xi) = u_{x\infty}
\tanh ( u_{x\infty} \xi )$. The propagation constants are respectively given by $\widetilde{\beta} = u_{x0}^2 /2$
and $\widetilde{\beta} = - u_{x\infty}^2$, which can also be found, {\it mutatis mutandis}, from Eqs.(16) and the
first of Eqs.(19), whenever the paraxial conditions ($u_{x0} << 1$ for bright solitons and $u_{x\infty} << 1$ for
dark solitons) are satisfied. These solitons obviously coincide with the asymptotic paraxial limit of the solitons
described in this paper. In order to make this comparison more quantitative, in Figure 3 and Figure 6 we have
superimposed to the exact soliton existence curves (solid curves) their paraxial counterparts (dashed curves).
More precisely, the FWHM of bright and dark paraxial solitons are easily shown to be $\widetilde{\Delta}_{bright}
= 2.6348 / u_{x0}$ and $\widetilde{\Delta}_{dark} = 1.0986 / u_{x\infty}$. As expected, the paraxial and exact
curves are practically indistinguishable for small $u_{x0}$ or $u_{x\infty}$. Not surprisingly, for dark solitons,
the agreement between exact and paraxial prediction is satisfactory almost everywhere since the value of
$u_{x\infty}$ is restricted to be less than $1/\sqrt{6}$ that is to a moderate nonparaxial regime.
\section{Two dimensional case: azimuthally polarized spatial dark solitons}
In order to deal with the two dimensional case, we introduce polar cylindrical coordinates $r,\varphi,z$ with unit
vectors $ \hat{\bf e}_r, \hat{\bf e}_\varphi,  \hat{\bf e}_z$ and look for fields of the form
\begin{equation} \label{field}
{\bf E}(r,\varphi\,z) = E_\varphi (r,z) \hat{\bf e}_\varphi + E_z(r,z) \hat{\bf e}_z,
\end{equation}
describing a circularly symmetric configuration with vanishing radial component. Inserting Eq.(\ref{field}) in
Eq.(\ref{maxwell2}), we obtain
\begin{eqnarray} \label{set}
&& \frac{\partial^2 E_z}{\partial r \partial z} = 0, \nonumber \\
&& \frac{\partial ^2 E_\varphi}{\partial z^2} + \frac{\partial}{\partial r} \left( \frac{\partial
E_\varphi}{\partial r} +\frac{E_\varphi}{r} \right) = -k^2 E_\varphi - k^2 \frac{4}{3} \frac{n_2}{n_0}  \left[
|{\bf E}|^2 E_\varphi + \frac{1}{2}
({\bf E \cdot E}) E_\varphi^* \right], \nonumber \\
&& \frac{\partial ^2 E_z}{\partial r^2} + \frac{1}{r} \frac{\partial E_z}{\partial r} = -k^2 E_z  - k^2
\frac{4}{3} \frac{n_2}{n_0}  \left[ |{\bf E}|^2 E_z + \frac{1}{2} ({\bf E \cdot E}) E_z^* \right].
\end{eqnarray}
Internal consistency of the set of Eqs.(\ref{set}) (three equations in two unknowns) requires $E_z = 0$. As a
consequence, the second of Eqs.(\ref{set}) yields
\begin{equation} \label{prop}
\frac{\partial ^2 E_\varphi}{\partial z^2} + \frac{\partial}{\partial r} \left( \frac{\partial E_\varphi}{\partial
r} +\frac{E_\varphi}{r} \right) =  -k^2 E_\varphi -2 k^2 \frac{n_2}{n_0}  |E_\varphi|^2 E_\varphi.
\end{equation}
We note that circular symmetry and polarization imposed to the field, together with the symmetry properties of
Kerr effect, have allowed us to reduce Maxwell's equations to the single Eq.(\ref{prop}). Equation (\ref{prop}) is
conveniently rewritten in the dimensionless form
\begin{equation} \label{adim}
\frac{\partial ^2 U}{\partial \zeta^2} + 2 \frac{\partial}{\partial \rho} \left( \frac{\partial U}{\partial \rho}
+\frac{U}{\rho} \right) =  -U - 2 \gamma |U|^2 U,
\end{equation}
where $\rho=\sqrt{2} k r$, $\zeta = kz$, $U =\sqrt{|n_2|/n_0} E_\varphi$ and $\gamma = n_2 / |n_2|$. If we look
for soliton solutions of the form
\begin{equation} \label{sol2}
U(\rho,\zeta) = e^{i \alpha \zeta} u(\rho),
\end{equation}
Eq.(\ref{adim}) becomes
\begin{equation} \label{eqsol}
\frac{d}{d \rho}\left( \frac{d u}{d \rho} + \frac{u}{\rho} \right) = \frac{1}{2}(\alpha^2-1) u - \gamma u ^3.
\end{equation}
Both the structure of Eq.(\ref{eqsol}) and the azimuthal field polarization dictate $u(0)=0$, so that azimuthally
polarized bright solitons do not exist. In order to find dark solitons, we introduce the further condition
\begin{equation}
\lim_{\rho \rightarrow \infty} u(\rho) = u_\infty,
\end{equation}
together with the vanishing of all derivatives for $\rho \rightarrow \infty$. Since focusing media ($\gamma = 1$,
i.e., $n_2>0$) are not able to support dark solitons, we consider hereafter defocusing media ($\gamma = -1$, i.e.,
$n_2<0$), so that Eq.(\ref{eqsol}) reads
\begin{equation} \label{eqdef}
\frac{d}{d \rho}\left( \frac{d u}{d \rho} + \frac{u}{\rho} \right) = \frac{1}{2}(\alpha^2-1) u + u ^3,
\end{equation}
which implies, together with the above boundary condition at infinity,
\begin{equation} \label{alf}
\alpha = \pm \sqrt{1-2u_\infty^2}.
\end{equation}
While positive and negative signs of $\alpha$ respectively refer to forward and backward travelling solitons (see
Eq.(\ref{sol2})), $u(\rho)$ depends on $\alpha^2$ (see Eq.(\ref{eqsol})). Equation (\ref{alf}) shows the existence
of an upper threshold for the soliton asymptotic amplitude
\begin{equation} \label{threshold}
u_\infty < \frac{1}{\sqrt{2}},
\end{equation}
since, otherwise, $\alpha$ would become imaginary. If we now insert Eq.(\ref{alf}) into Eq.(\ref{eqdef}), we
obtain
\begin{equation} \label{eqsol1}
\frac{d}{d \rho}\left( \frac{d u}{d \rho} + \frac{u}{\rho} \right) = (u^2-u_\infty^2) u.
\end{equation}

We have carried out a numerical integration of Eq.(\ref{eqsol1}) with boundary conditions $u(0)=0$ and
$u(\infty)=u_\infty$, by employing a standard shooting-relaxation method for boundary value problems. The results
of our simulations show that dark solitons can be obtained in the range of field amplitudes $0 < u_\infty <
1/\sqrt{2}$. Different soliton profiles are reported in Figure 8.

In order to complete our analysis, we now evaluate both the magnetic field and the Poynting vector. Recalling the
expression of the soliton electric field
\begin{equation} \label{ElecField}
{\bf E} = \sqrt{\frac{n_0}{|n_2|}} e^{i \alpha k z } u(\sqrt{2}kr)  \hat{\bf e}_\varphi,
\end{equation}
we obtain, from the first of Eqs.(\ref{maxwell1}) written in cylindrical coordinates,
\begin{equation} \label{MagField}
{\bf B} = - \sqrt{\frac{n_0}{|n_2|}} e^{i \alpha k z } \frac{k}{\omega}
         \left[ \alpha u \hat{\bf e}_r + i \sqrt{2} \left( \frac{du}{d\rho} + \frac{u}{\rho} \right)
         \hat{\bf e}_z \right]_{\rho=\sqrt{2}kr}.
\end{equation}
The magnetic field has a radial component whose shape coincides with that of the electric field, and a vanishing
azimuthal component, so that $\bf E$ and $\bf B$ are mutually orthogonal. With the help of Eqs.(\ref{ElecField})
and (\ref{MagField}), the time averaged Poynting vector
\begin{equation} \label{Poy}
{\bf S} = \frac{1}{2 \mu_0} Re \left( {\bf E} \times {\bf B}^* \right)
\end{equation}
turns out to be given by
\begin{equation} \label{Poyn}
{\bf S} (r) =  \frac{\alpha k}{2 \omega \mu_0} \frac{n_0}{|n_2|} u^2(\sqrt{2}kr) \hat{\bf e}_z = \frac{\alpha k}{2
\omega \mu_0} |{\bf E}|^2 \hat{\bf e}_z.
\end{equation}
We note that $\bf S$ is parallel to the $z-$ axis, consistently with the shape-invariant nature of solitons. From
an analytical point of view, this corresponds to the $\pi /2$ phase difference between $B_z$ and $E_\varphi$ (see
Eqs.(\ref{ElecField}) and (\ref{MagField})). As expected, the Poynting vector is either parallel or antiparallel
to $\hat{\bf e}_z$ according to the sign of $\alpha$, while its amplitude is proportional to $|{\bf E}|^2$. The
above plane wave-like properties are consistent with the nondiffractive nature of exact solitons.

It is worthwhile to underline that, in the case of the azimuthal dark solitons we are considering, the asymptotic
optical intensity $I_\infty = |{\bf S(\infty)}|$ turns out not to be proportional to $u_\infty^2$. In fact, by
using Eqs.(\ref{alf}) and (\ref{Poyn}), one obtains
\begin{equation} \label{Inw}
I_\infty(u_\infty) = I_0 u_\infty^2 \sqrt{1-2u_\infty^2}
\end{equation}
where $I_0 = kn_0 /(2\omega \mu_0 |n_2|)$, whose profile is reported in Figure 9. Equation (\ref{Inw}) shows that
the asymptotic optical intensity is not a monotonically increasing function of the asymptotic field amplitude, but
reaches its maximum threshold value $I_\infty^{max} = I_0 / 3^{3/2}$ in correspondence to $u_\infty = 1/\sqrt{3}$.
This is connected to the $\alpha-$ dependence of the magnetic field (see Eq.(\ref{MagField})) whose radial part
tends to vanish for $u_\infty \rightarrow 1/\sqrt{2}$. A related and relevant consequence of Eq.(\ref{Inw}) is the
existence of {\it two solitons} of different widths for a given asymptotic optical intensity. The existence curve
relating the normalized half width at half maximum (HWHM) of the soliton optical intensity profile $|{\bf
S}(\rho)|$ to $u_\infty$ is reported in Fig.10. In particular, Fig.10 shows the existence of a normalized minimum
HWHM $\simeq 2.1$ ($\simeq 0.24 \lambda$) for $u_\infty = 1/\sqrt{2}$. In addition, Fig.9 shows that a normalized
HWHM $\simeq 2.7$ ($\simeq 0.3 \lambda$) corresponds to $u_\infty = 1/\sqrt{3}$ for which the soliton attains the
maximum asymptotic optical intensity $I_\infty^{max}$.

It is interesting to examine the behavior of our solution in the limit of large $\rho$. To this end, neglecting in
Eq.(\ref{eqsol1}) the term in $u/\rho$, we have
\begin{equation} \label{nls}
\frac{d^2 u}{d \rho^2} = (u^2 - u_\infty^2) u
\end{equation}
which formally coincides with the equation describing one dimensional linearly polarized paraxial dark solitons.
Equation (\ref{nls}) admits of the solution $u= u_\infty \tanh (\rho u_\infty /\sqrt{2})$. This solution can be
compared with the exact one. This is done in Figure 11 where the ratio $R(\rho)$ between the hyperbolic tangent
and the exact solution is reported as function of $\rho$, for different values of $u_\infty$. The hyperbolic
tangent solution reproduces the exact one for large values of $\rho$, as expected, while it at most differs by a
factor $\cong 1.2$ for small values of $\rho$.
\section{Conclusions}
In this paper, the problem of the existence of nonparaxial spatial Kerr solitons has been completely solved. This
has been done by showing that spatial solitons can be derived as exact solutions of Maxwell equations (thus
making, within our approach, the term "paraxial"  and "nonparaxial" redundant). In the one dimensional case, the
{\it perfect optical soliton} represents the straightforward generalization of the paraxial one, the main
difference being that dark solitons exhibit, unlike their paraxial counterpart, a specific upper limit for the
possible values their asymptotic intensity can assume. In the two dimensional case, the exact dark soliton is of a
completely new kind, and the difference between paraxial and nonparaxial becomes rather meaningless. In any case,
the comparison between paraxial and exact solitons, done, for example, by inspecting the relative existence
curves, shows that our solitons are a definite entity, independent from used the approximation scheme: the
transition between the paraxial and the highly diffractive regime is very smooth and does not exhibit any kind of
dramatic catastrophic behavior, as implied by the standard paraxial theory.
\appendix
\section{Appendix A: Existence of bright solitons in focusing media}
In order to tackle the problem of bright solitons existence we have to prove that the curve defined in
Eq.(\ref{BriCon}), with $\beta^2$ given by Eq.(\ref{betasq}), actually reaches the origin of the phase plane
$(u_x,u_z)$. To this end, it is convenient to introduce the polar coordinate defined by $u_x = \rho \cos \phi$ and
$u_z = \rho \sin \phi$, so that Eq.(\ref{BriCon}) becomes
\begin{eqnarray}
\rho^2 \left\{ \rho^4 \cos^2 \phi \left[ 2 \cos^2 \phi + \frac{4}{3} \cos^2\phi \sin^2 \phi + \frac{2}{9} \sin^4
\phi \right] \right. \nonumber \\
+ \gamma \rho^2 \left[ \frac{1}{2}(4-3\beta^2) \cos^4 \phi + \frac{1}{3}(2-\beta^2) \cos^2 \phi \sin^2 \phi +
\frac{1}{2} \beta^2 \sin^4 \phi \right] \nonumber \\
\left. + \frac{1}{2} \left[ (1-\beta^2) \cos^2 \phi + \beta^2 \sin^2 \phi \right] \right\} = 0.
\end{eqnarray}
This equation is trivially satisfied by setting $\rho=0$ and this is consistent with the fact that the origin is
in itself an orbit of Eqs.(\ref{system2}). Therefore, the integral curve associated with solitons is described by
the vanishing of the expression within the curly brackets. Requiring that this curve reaches the origin yields
\begin{equation} \label{compbri}
\tan^2 \phi_0 = 1 - \frac{1}{\beta ^2} \equiv u_{0x}^2 \frac{4 u_{0x}^2 + \gamma}{(1 + 2 \gamma u_{x0}^2)^2}
\end{equation}
where $\phi_0 = \phi(\rho=0)$ and $\beta^2$ has been obtained from Eq.(\ref{betasq}). For $\gamma = 1$, the RHS of
Eq.(\ref{compbri}) is positive so that this equation can always be solved which, together with the fact that
Eq.(\ref{noncrossellipse}) is always satisfied for $\gamma =1$, implies that bright soliton exist in focusing
media for any value of $u_{0x}$. In the case $\gamma = -1$, the RHS of Eq.(\ref{compbri}) is positive for
$|u_{0x}| > 1/2$ so that, the curve actually reaches the origin. However, in this case, the curve joining the
points $(u_x,u_z) = (0,0)$ and $(u_x,u_z) = (u_{x0},0)$ unavoidably crosses the ellipse $ 6 u_x^2 + (2/3) u_z^2  =
1$ since its semi-axis along the $x-$axis is $1/\sqrt{6} < 1/2$. Therefore, for $\gamma = -1$, a point belonging
to the integral curve such that Eq.(\ref{noncrossellipse}) fails to be satisfied always exists, with the
consequence that, in defocusing media, bright solitons never exist.
\section{Appendix B: Conditions for dark solitons existence}
As already explained in Section 2, Eqs.(\ref{q1}) and (\ref{q2}) are necessary for dark soliton existence so that
we have to find when they are also sufficient. From Eq.(\ref{q1}) (with $\gamma =-1$) it is evident that solitons
can exist for $u_{x\infty}^2 < 1/2$. The equation for the dark soliton integral curve on the phase plane
(Eq.(\ref{levelset}) with $\beta$ and $F_0$ given in Eqs.(\ref{darkexist})) can be solved for $u_z^2$ thus
yielding
\begin{eqnarray} \label{uzsq}
u_z^2 =- \frac{\frac{8}{3} u_x^4 - \frac{2}{3} (1+2u_{x\infty}^2)u_x^2 + (1-2u_{x\infty}^2)}
         {\frac{8}{9}u_x^2 - 2(1-2u_{x\infty}^2)} \nonumber \\
      + \frac{\sqrt{\left[\frac{8}{3} u_x^4 - \frac{2}{3} (1+2u_{x\infty}^2)u_x^2 + (1-2u_{x\infty}^2)\right]^2
         - \left[\frac{16}{9}u_x^2 - 4(1-2u_{x\infty}^2)\right]
         \left[ 4u_x^6 -(1+6u_{x\infty}^2)u_x^4 + 2u_{x\infty}^2 u_x^2 - (1-2 u_{x\infty}^2)u_{x\infty}^4 \right] }}
         {\frac{8}{9}u_x^2 - 2(1-2u_{x\infty}^2)} \nonumber \\
\end{eqnarray}
which furnishes $u_z^2$ as a function of $u_x^2$ (parametrically dependent on $u_{x\infty}^2$) along the dark
soliton integral curve. Here, the plus sign between the two terms has been chosen in order to satisfy the boundary
condition $u_z (u_{x\infty}^2) = 0$. Evaluating Eq.(\ref{uzsq}) at $u_x = 0$ and taking the square root of the
result we obtain the second of Eqs.(\ref{darkexist}), which is consistent with the boundary conditions since it
satisfies the second of Eqs.(\ref{q2}) (with $\gamma = -1$). Therefore, in order to prove soliton existence, we
are left with proving that the curves in Eq.(\ref{uzsq}) actually reach the point ${\bf u}_\infty = (u_{x\infty}
\: \: 0)^T$ (i.e., with proving that the RHS of Eq.(\ref{uzsq}) is a positive real number) and that such curves do
not cross the ellipse $ 6 u_x^2 + (2/3) u_z^2 = 1$, thus leaving Eq.(\ref{noncrossellipse}) satisfied. Since the
expression under the square root is always positive for $u_{x\infty}^2<1/2$, we have only to ensure that the RHS
of Eq.(\ref{uzsq}) is positive. It is not difficult to show that this is the case whenever
\begin{equation}
4u_x^6 -(1+6u_{x\infty}^2)u_x^4 + 2u_{x\infty}^2 u_x^2 - (1-2 u_{x\infty}^2)u_{x\infty}^4 <0.
\end{equation}
Imposing that the maximum of the polynomial in the LHS of this inequality is negative, we obtain the condition
$u_{x\infty}^2< 1/6$. The existence of dark solitons in this range for $u_{x\infty}^2$ is finally proved by noting
that any integral curve associated to these solitons globally lies within the ellipse $ 6 u_x^2 + (2/3) u_z^2 =
1$, so that Eq.(\ref{noncrossellipse}) is always satisfied.

{\bf ACKNOWLEDGMENTS}

This research has been funded by the Istituto Nazionale di Fisica della Materia through the "Solitons embedded in
holograms", the FIRB "Space-Time nonlinear effects" projects and the Air Force Office of Scientific Research (H.
Schlossberg).

\clearpage {\Large \bfseries Figure Captions}
\begin{itemize}
   \item {\bfseries Figure 1}: Plot of phase portrait of Eqs.(\ref{system2}) associated to bright solitons for
   $|u_{x0}| = 1,2,3,4$. Each bow-tie shaped curve is obtained by plotting the level set defined in Eq.(\ref{BriCon})
   with $\beta$ given by Eq.(\ref{betabright}). Any piece of curve starting from and ending into the origin (left or
   right loop of each bow-tie) is associated to a single bright soliton.

   \item {\bfseries Figure 2}: Plot of the transverse component $u_x(\xi)$ (a) and longitudinal component
   $u_z(\xi)$ (b) of bright solitons for $u_{x0} = 1,2,3,4$ (same cases as in Figure 1) and $\beta > 0$.

   \item {\bfseries Figure 3}: Bright soliton existence curve (solid line), relating the FWHM, $\Delta_{bright}$,
   of the amplitude $u_x(\xi$) to $|u_{x0}|$. For very small and very large $|u_{x0}|$, the FWHM diverges
   and vanishes, respectively. The dashed line represents the FWHM, $\widetilde{\Delta}_{bright}$, of paraxial bright
   solitons. Note the complete overlapping of the two curves for $u_{x0} < 0.2$.

   \item {\bfseries Figure 4}: Plot of phase portrait of Eqs.(\ref{system2}) associated to dark solitons for
   $|u_{x\infty}| = 0.1,0.2,0.3,0.4$.  Each loop is obtained by plotting the level set defined in Eq.(\ref{levelset})
   with $\beta$ and $F_0$ given in Eqs.(\ref{darkexist}). Any piece of curve joining the points $(-u_{x\infty},0)$
   and $(u_{x\infty},0)$ is associated to a single dark soliton.

   \item {\bfseries Figure 5}: Plot of the transverse component $u_x(\xi)$ (a) and longitudinal component
   $u_z(\xi)$ (b) of dark soltions for $u_{x\infty} = 0.1,0.2,0.3,0.4$ (same cases as in Figure 4) and $\beta > 0$.

   \item {\bfseries Figure 6}: Dark soliton existence curve (solid line), relating the FWHM, $\Delta_{dark}$ of the
   amplitude $u_x (\xi)$ to $|u_{x\infty}|$. For very small $|u_{x\infty}|$, the FWHM diverges whereas at the
   threshold value $u_{x\infty} = 1/\sqrt{6}$ it attains its minimum value $\simeq 3$. The dashed line represents the
   FWHM, $\widetilde{\Delta}_{dark}$, of paraxial dark solitons. Note the complete overlap for most of the values of
   $u_{x\infty}$.

   \item {\bfseries Figure 7}: Normalized optical intensity $|{\bf S} (\xi)| / I_0$ of bright (a) and dark (b) solitons
   evaluated from Eq.(\ref{avpoy}) for the same soliton conditions as in Figure 1 (for bright solitons) and Figure 5
   (for dark solitons).

   \item {\bfseries Figure 8}: Two dimensional dark soliton profile $u(\rho)$ for various values of $u_\infty$.

   \item {\bfseries Figure 9}: Normalized asymptotic optical intensity $I_\infty / I_0$ as a function of the
   asymptotic dimensionless field amplitude $u_\infty$. Note that two solitons exist for any allowed asymptotic
   optical intensity.

   \item {\bfseries Figure 10}: Existence curve relating the normalized soliton optical intensity HWHM to $u_\infty$.

   \item {\bfseries Figure 11}: Plot of the ratio $R(\rho) = u_\infty \tanh (u_\infty \rho/\sqrt{2}) / u(\rho)$ for
   different values of $u_\infty$.
\end{itemize}

\clearpage
\begin{figure}
\label{F1} \clearpage \centering
\includegraphics[width=0.70\textwidth]{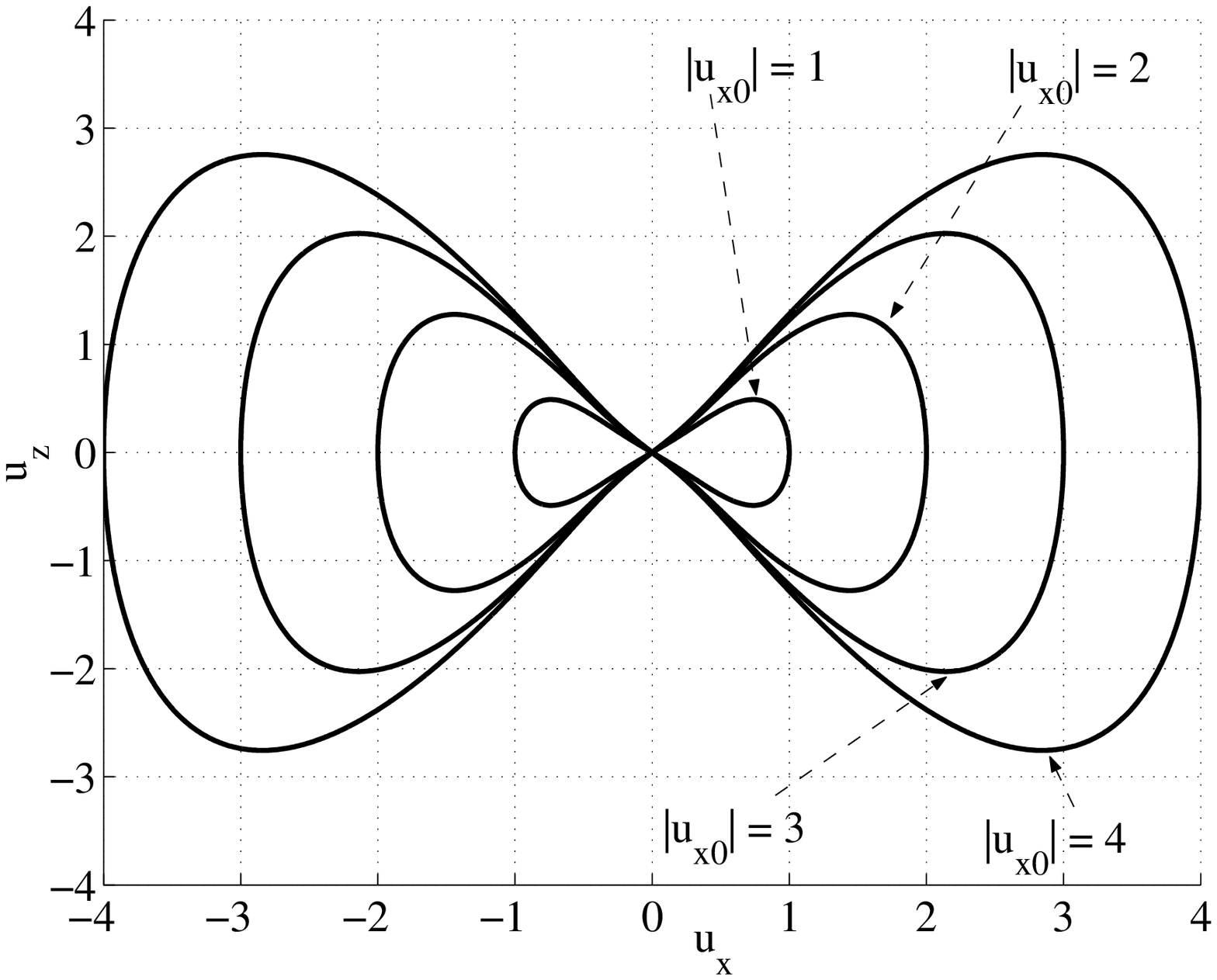}
\caption{}
\end{figure}
\begin{figure}
\label{F2} \clearpage \centering
\includegraphics[width=0.70\textwidth]{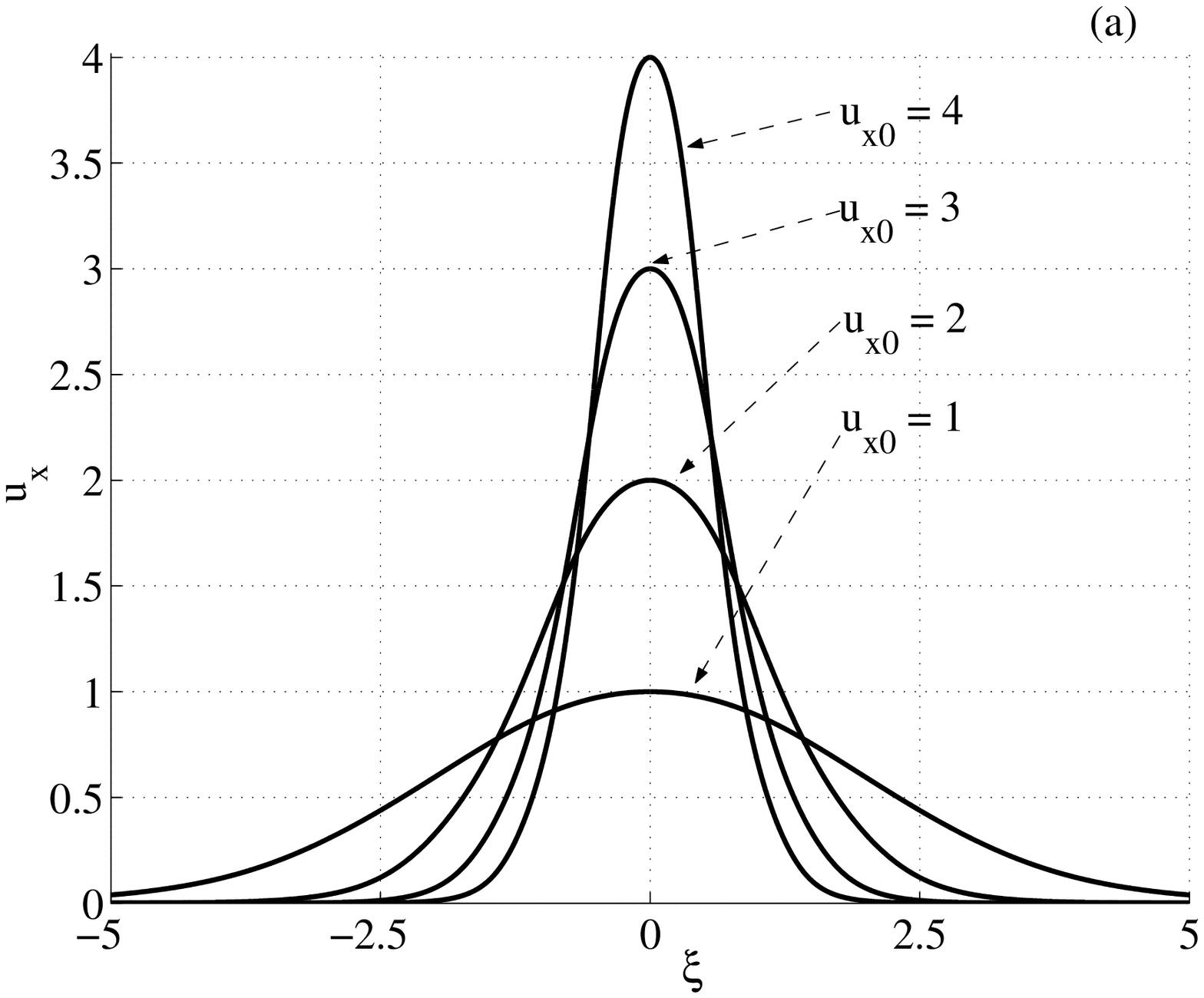}
\includegraphics[width=0.70\textwidth]{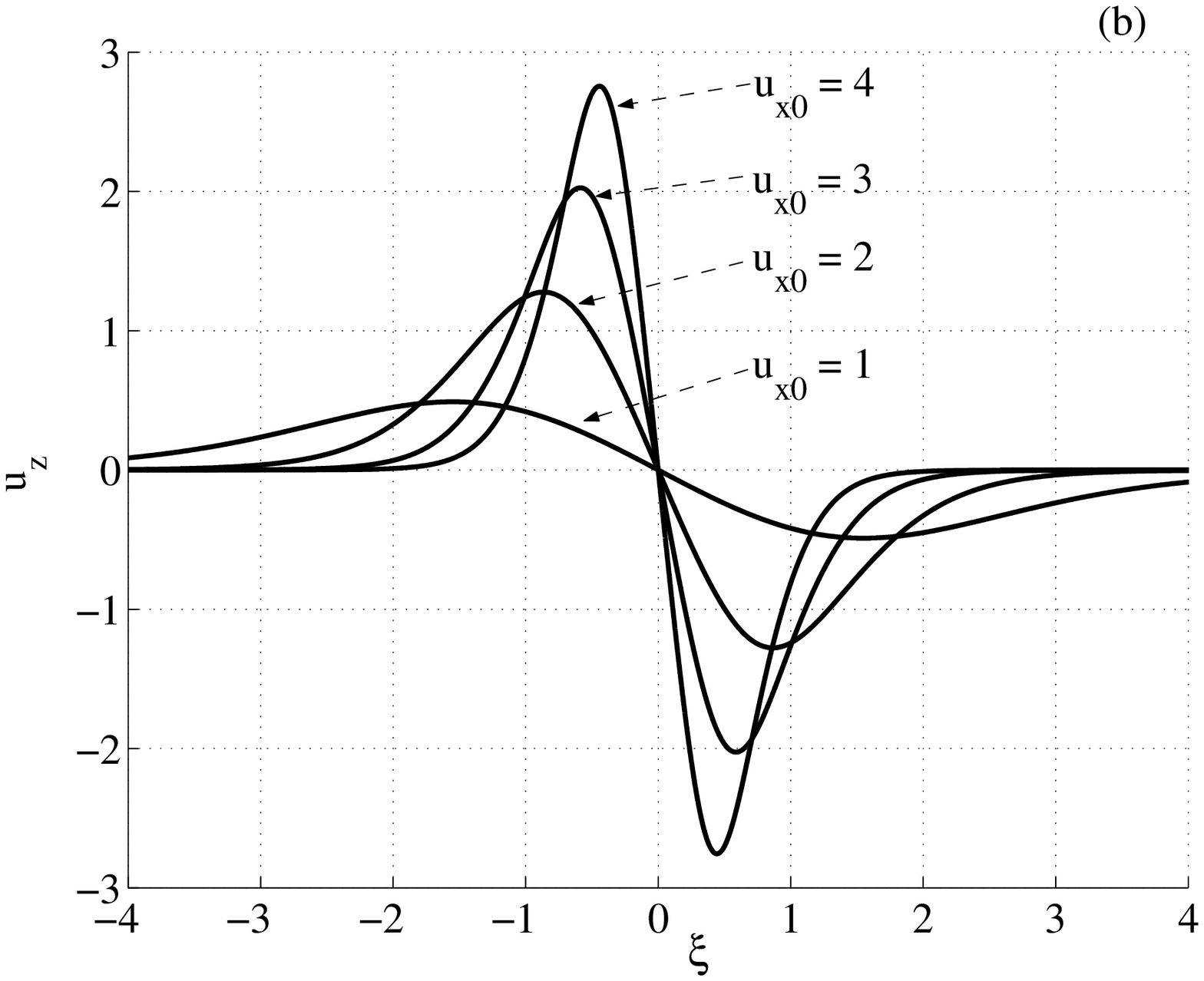}
\caption{}
\end{figure}
\begin{figure}
\label{F3} \clearpage \centering
\includegraphics[width=0.70\textwidth]{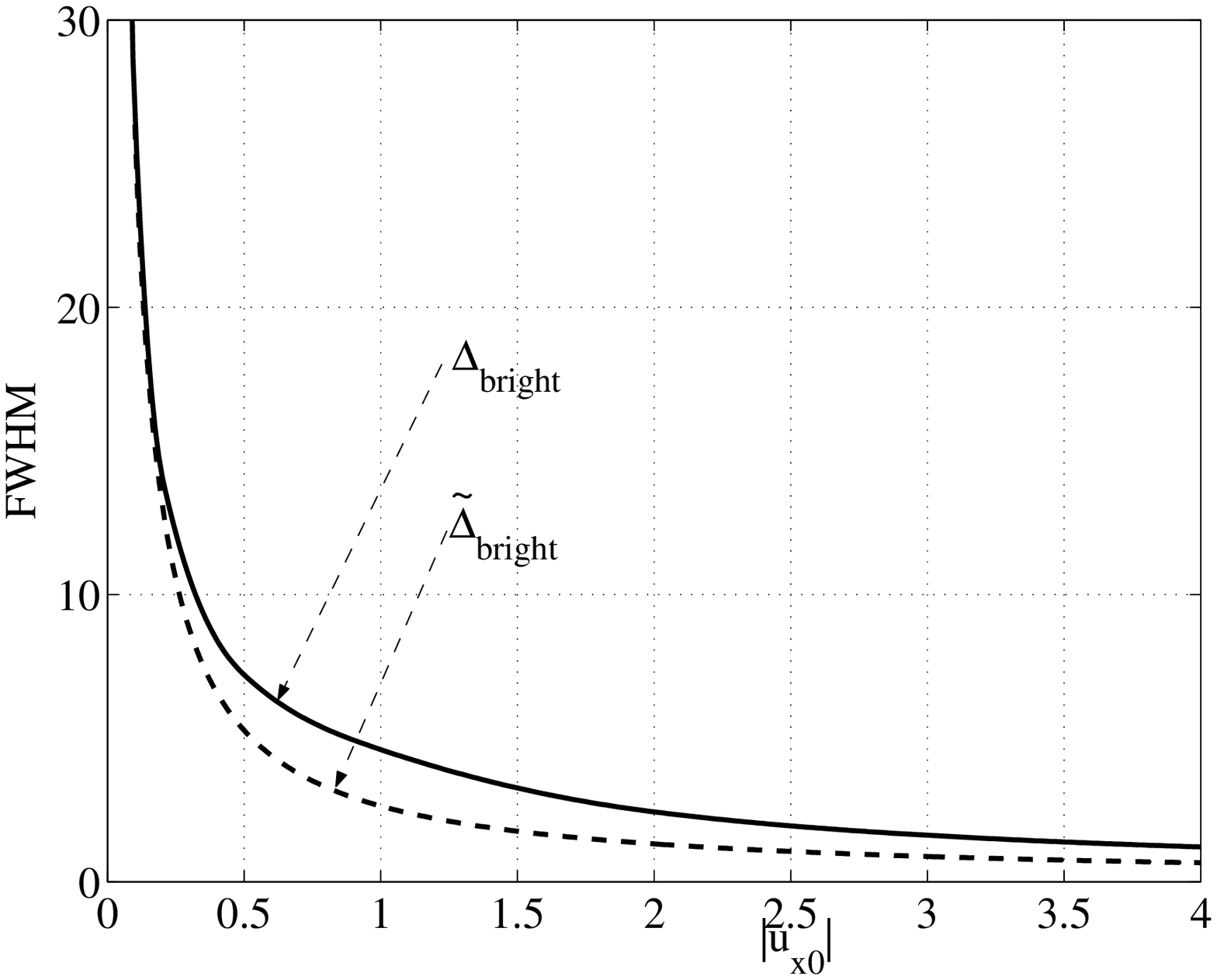}
\caption{}
\end{figure}
\begin{figure}
\label{F4} \clearpage \centering
\includegraphics[width=0.70\textwidth]{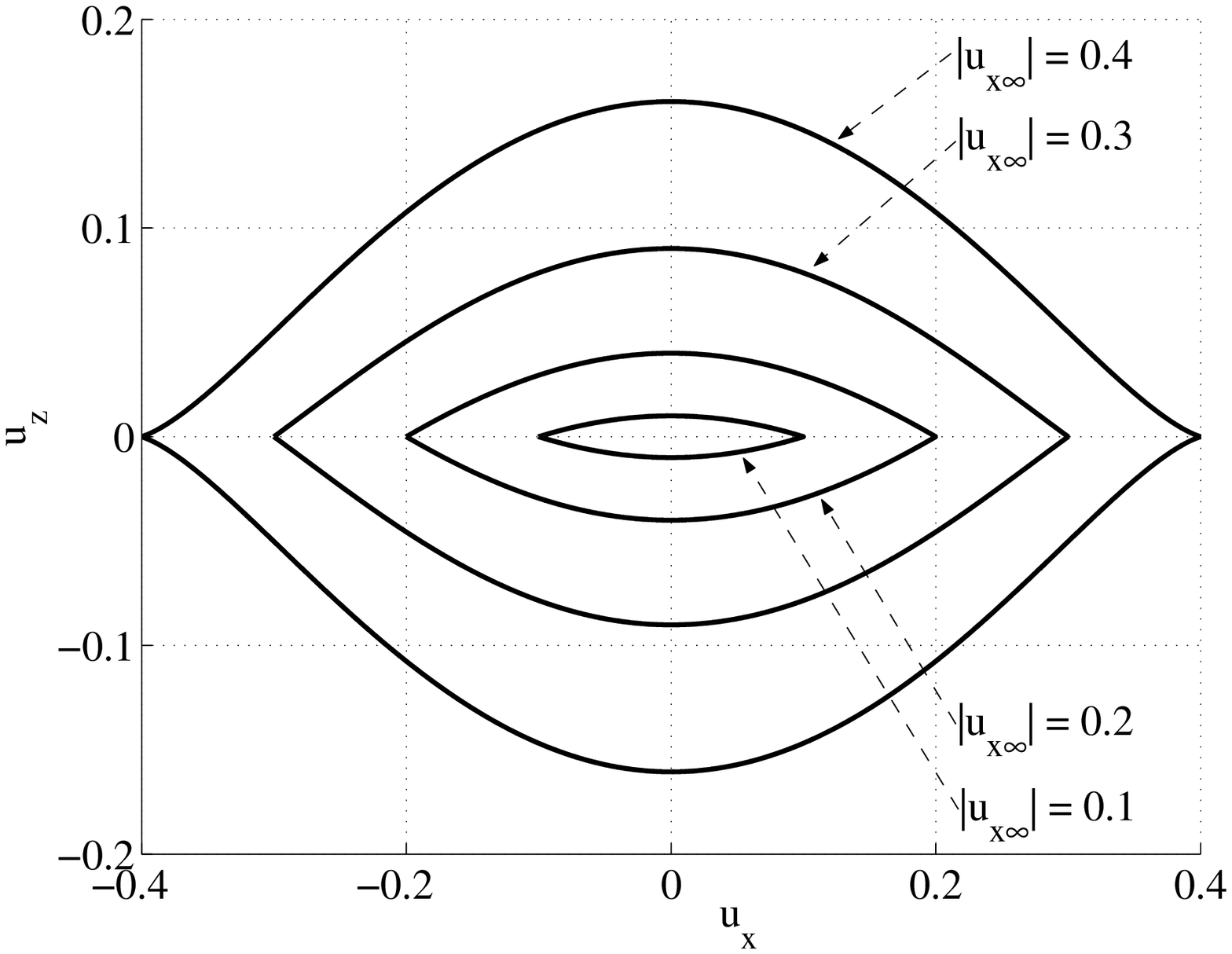}
\caption{}
\end{figure}
\begin{figure}
\label{F5} \clearpage \centering
\includegraphics[width=0.70\textwidth]{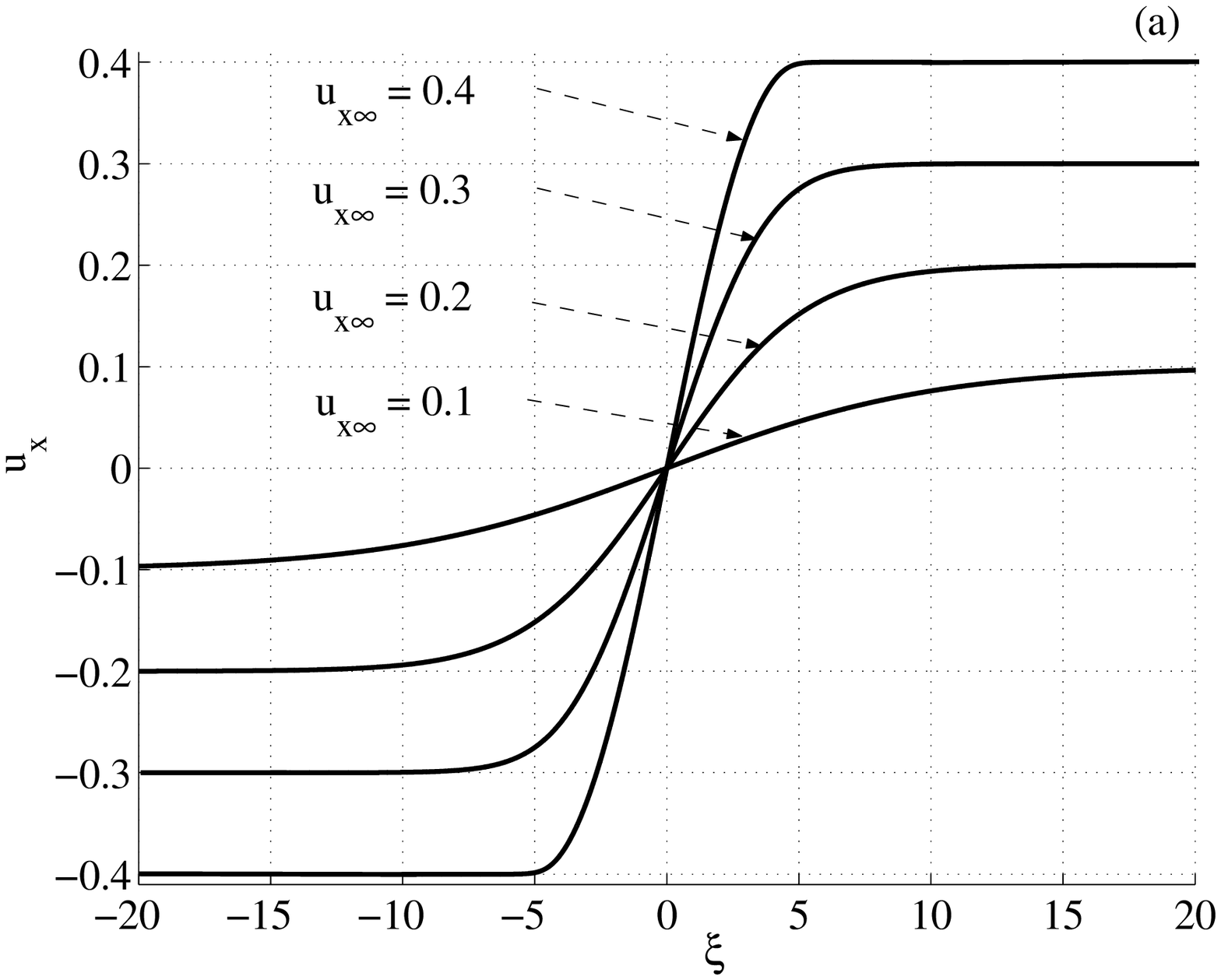}
\includegraphics[width=0.70\textwidth]{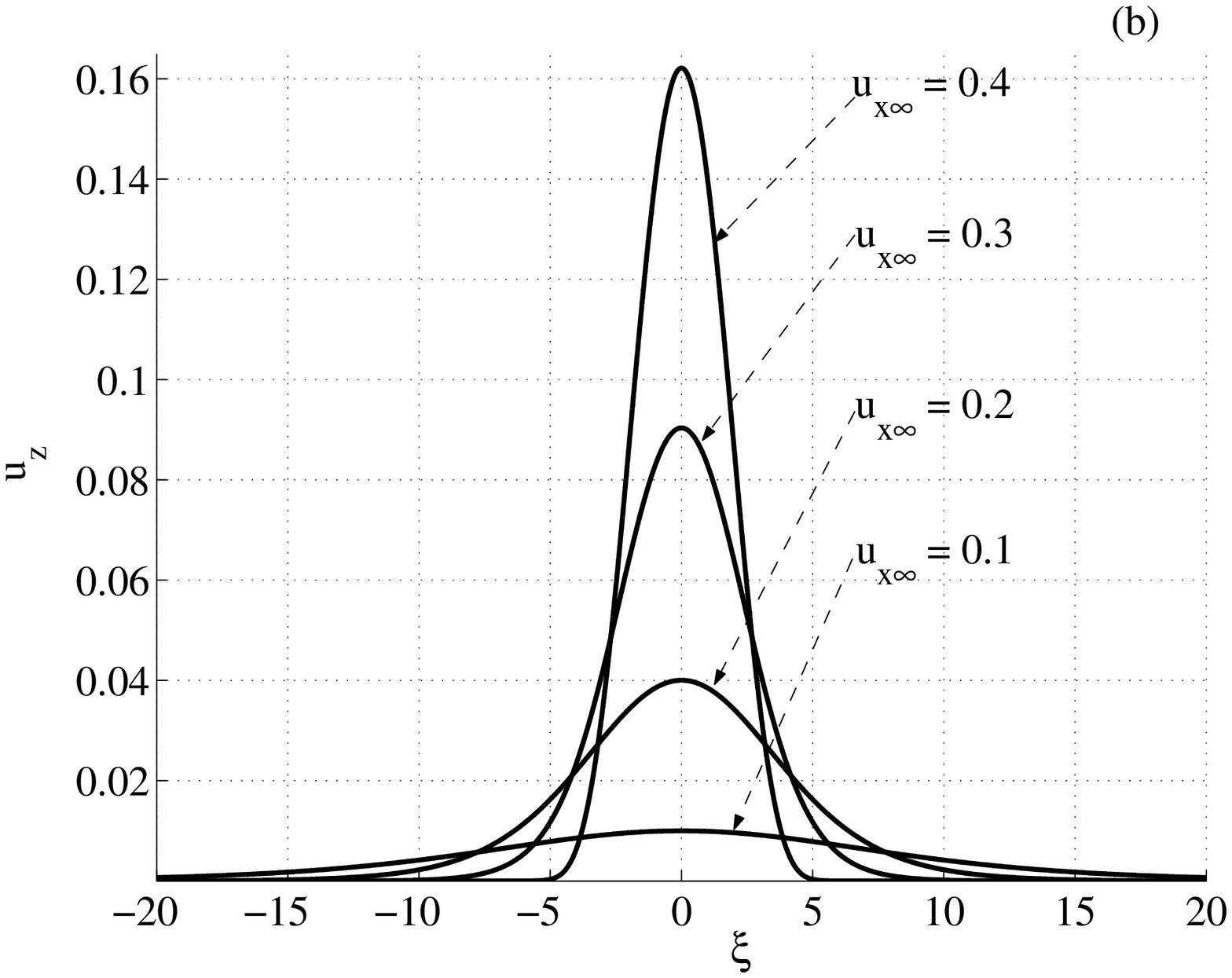}
\caption{}
\end{figure}
\begin{figure}
\label{F6} \clearpage \centering
\includegraphics[width=0.70\textwidth]{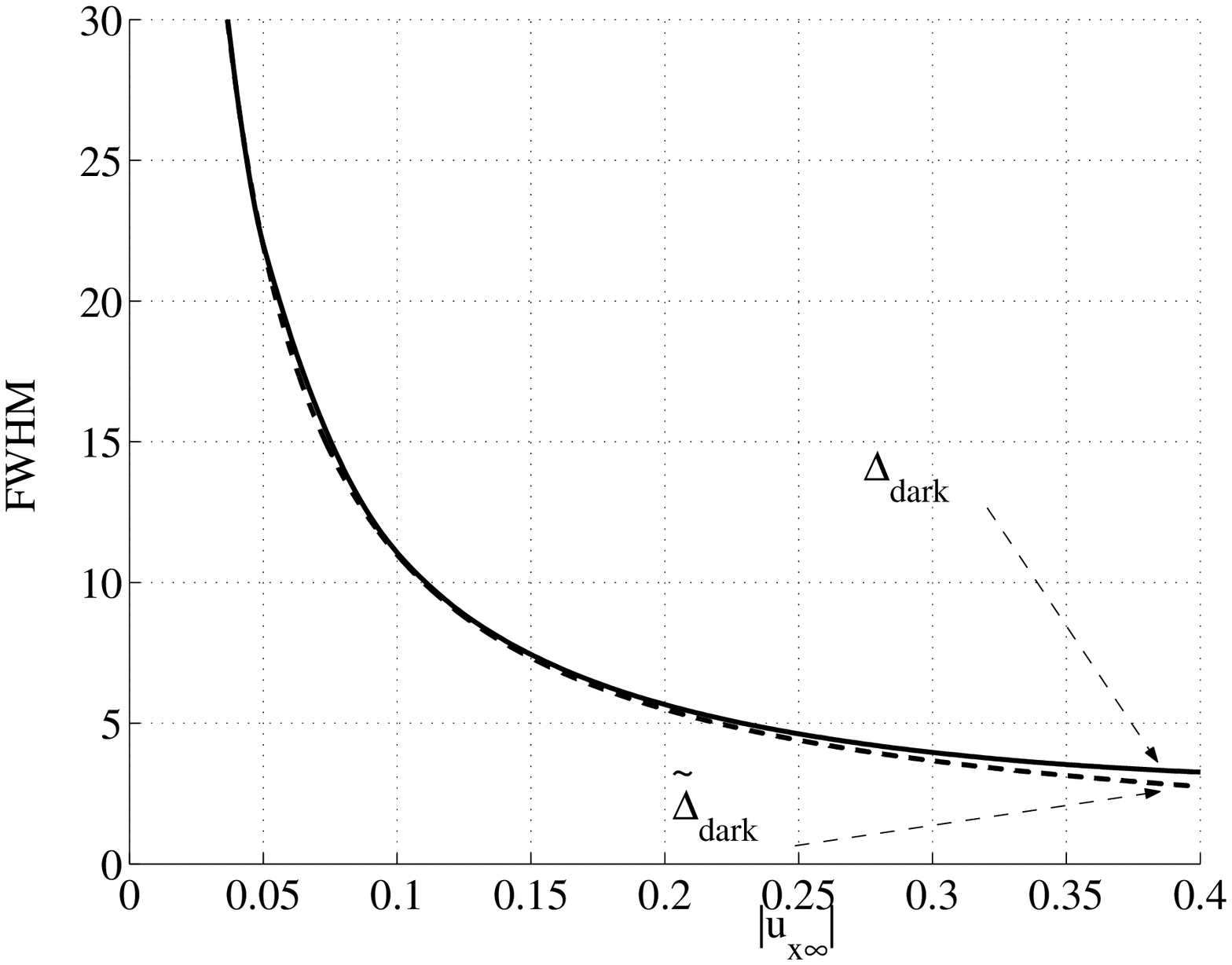}
\caption{}
\end{figure}
\begin{figure}
\label{F7} \clearpage \centering
\includegraphics[width=0.70\textwidth]{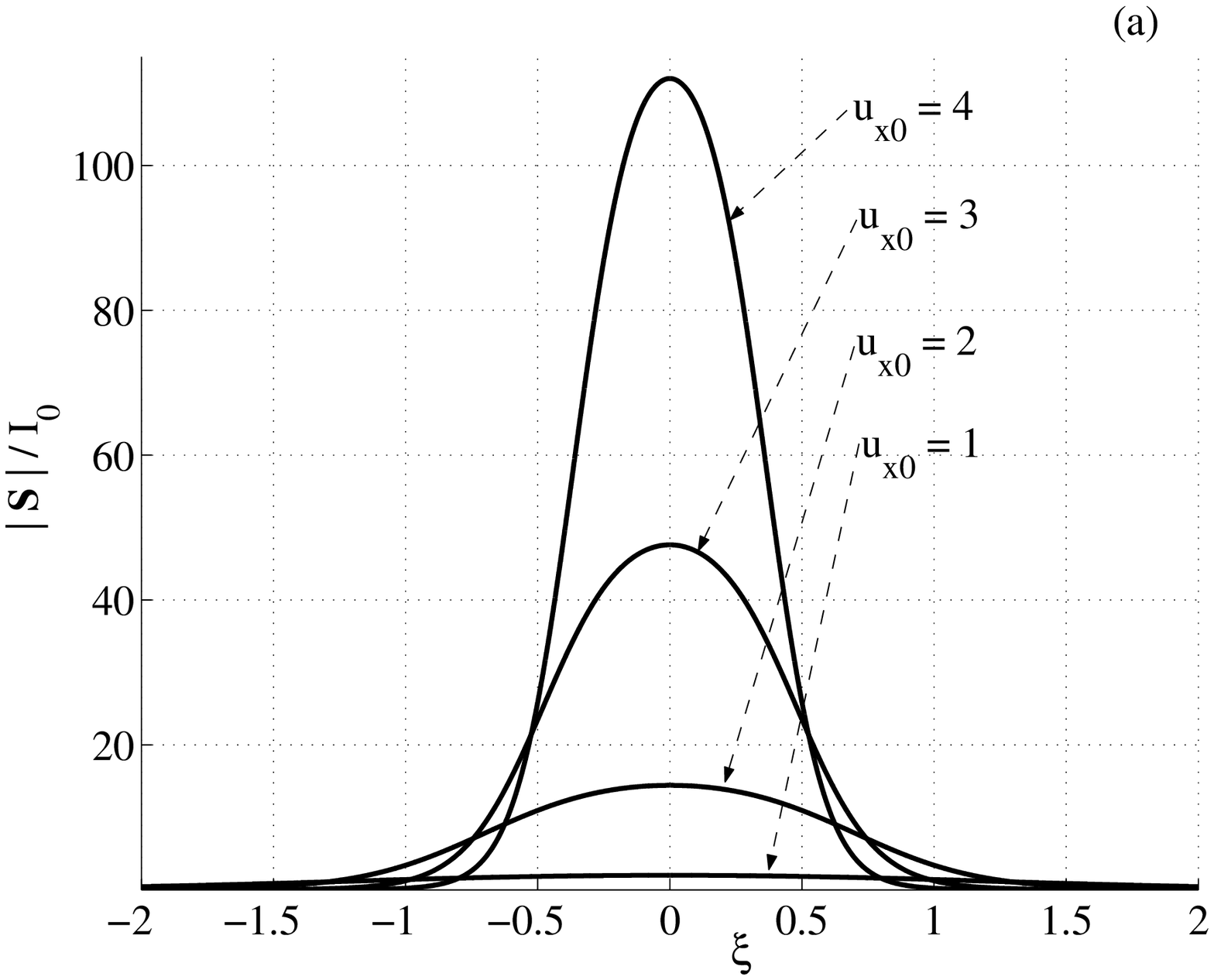}
\includegraphics[width=0.70\textwidth]{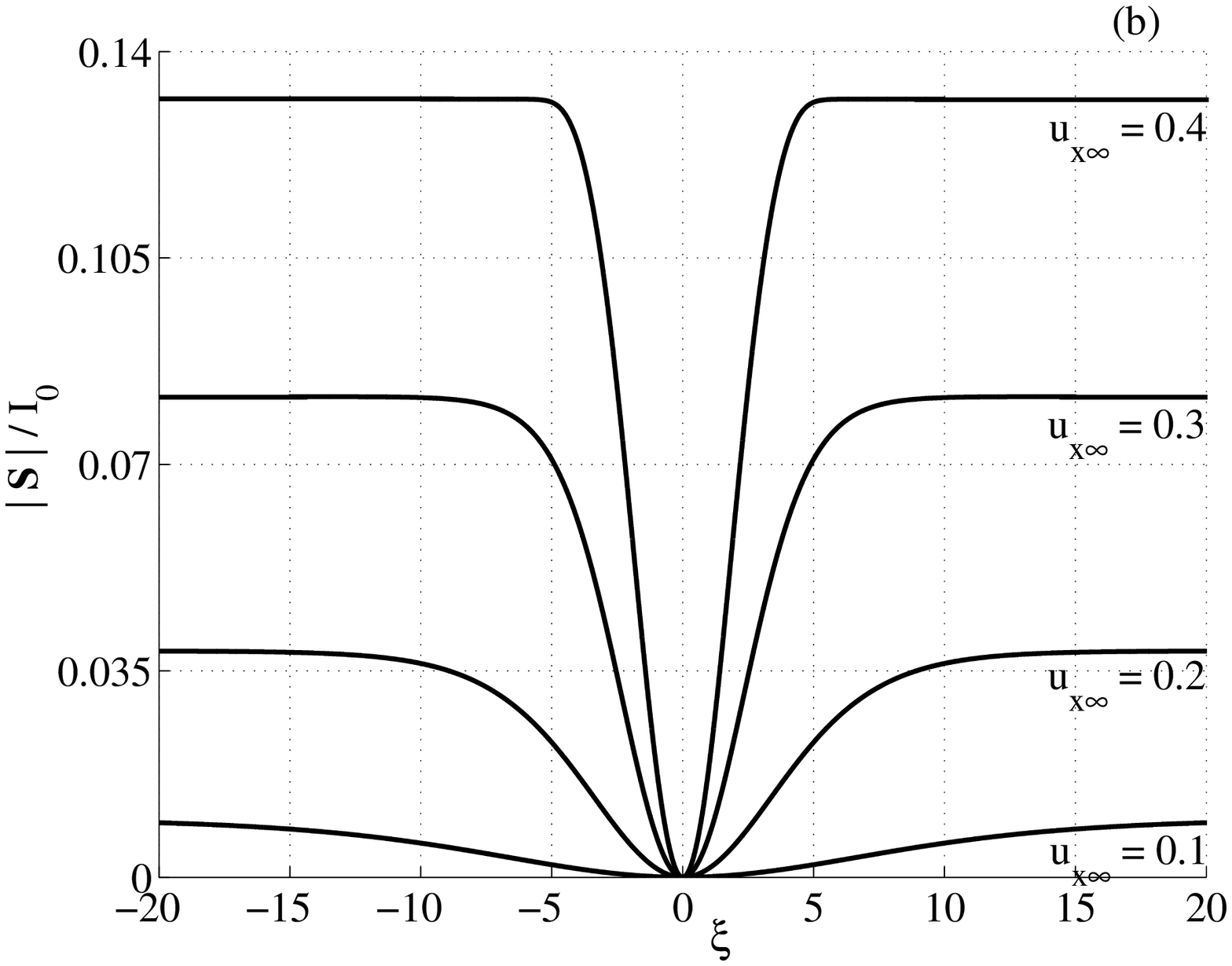}
\caption{}
\end{figure}
\begin{figure}
\label{F8} \clearpage \centering
\includegraphics[width=0.7\textwidth]{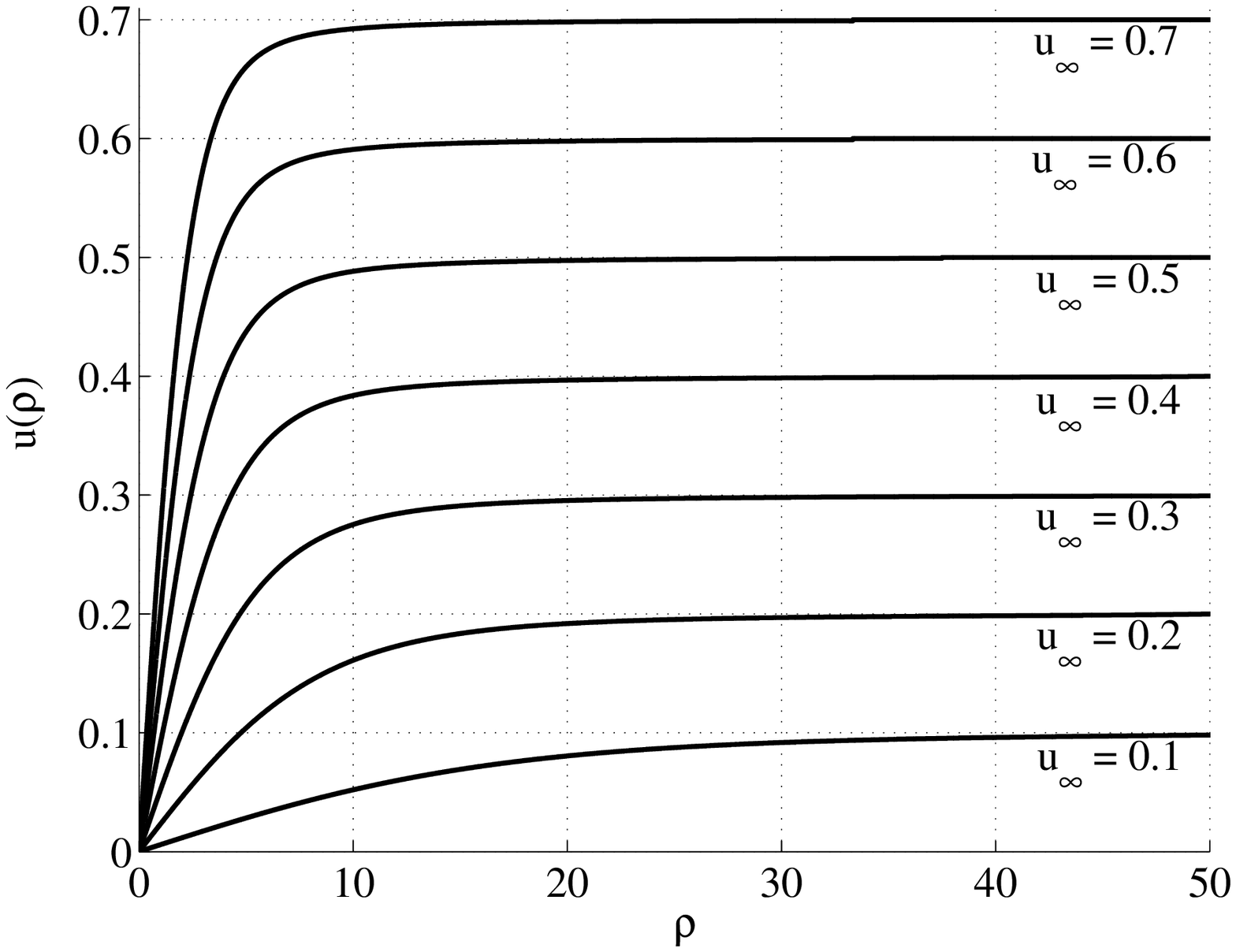}
\caption{}
\end{figure}
\begin{figure}
\label{F9} \clearpage \centering
\includegraphics[width=0.7\textwidth]{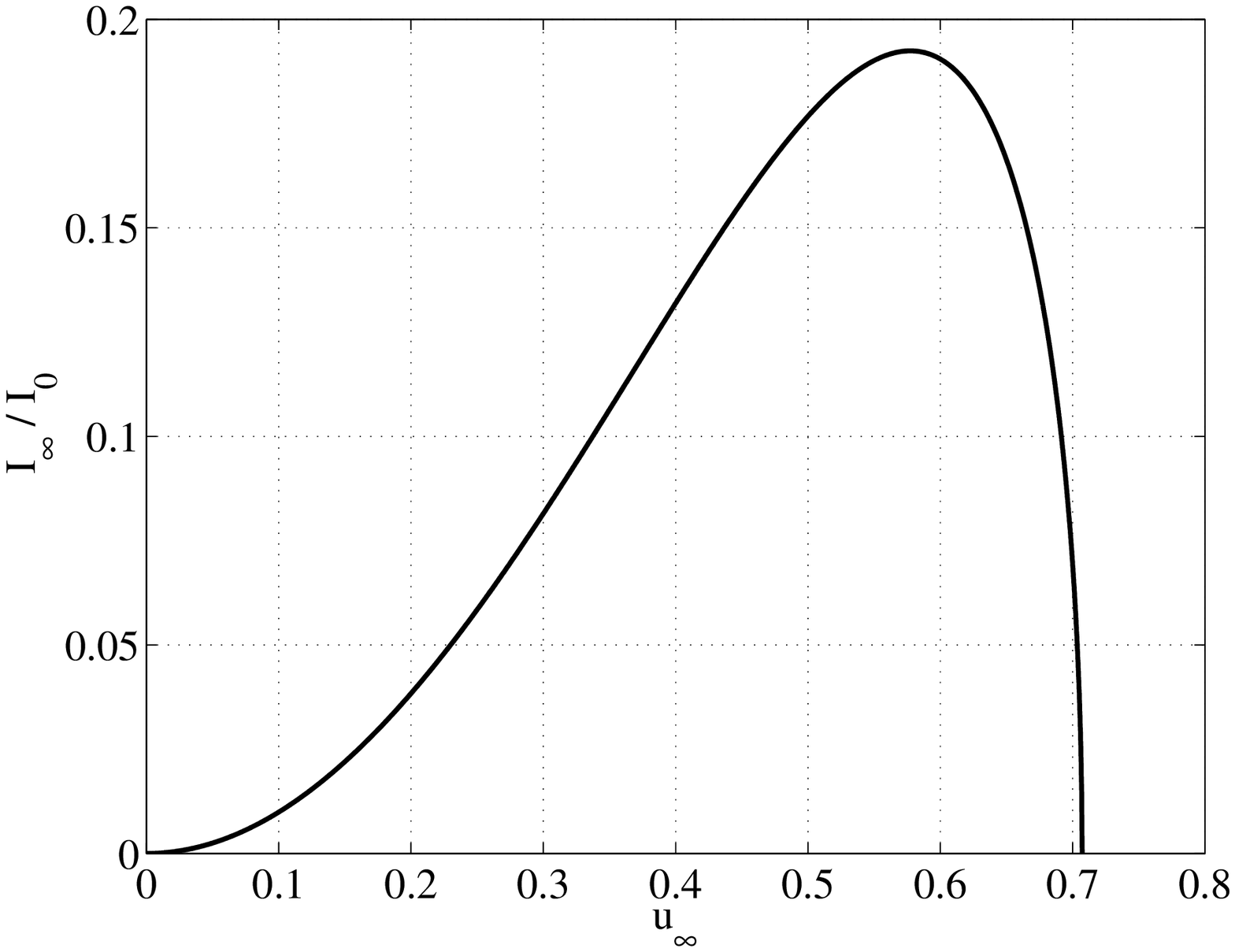}
\caption{}
\end{figure}
\begin{figure}
\label{F10} \clearpage \centering
\includegraphics[width=0.7\textwidth]{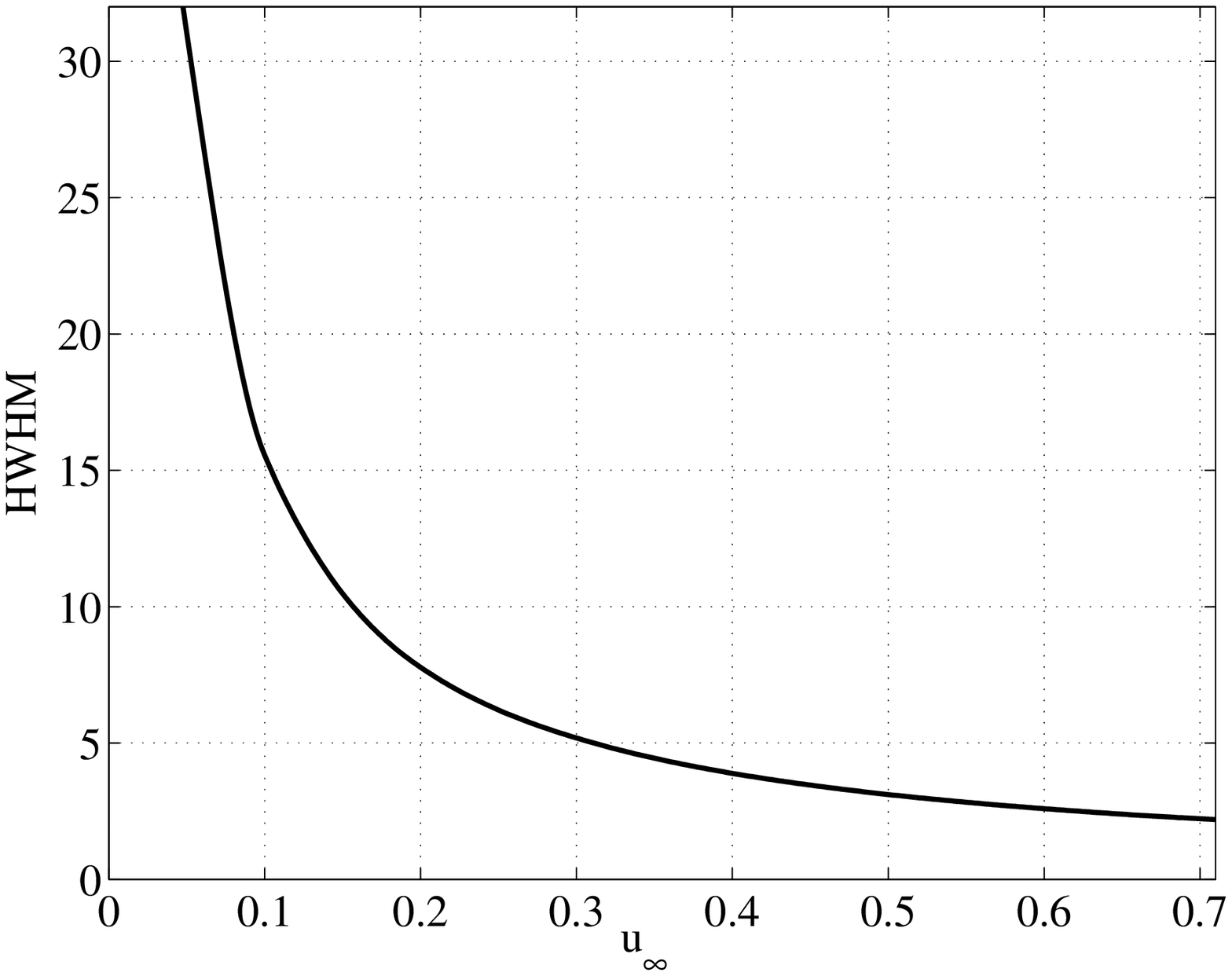}
\caption{}
\end{figure}
\begin{figure}
\label{F11} \clearpage \centering
\includegraphics[width=0.7\textwidth]{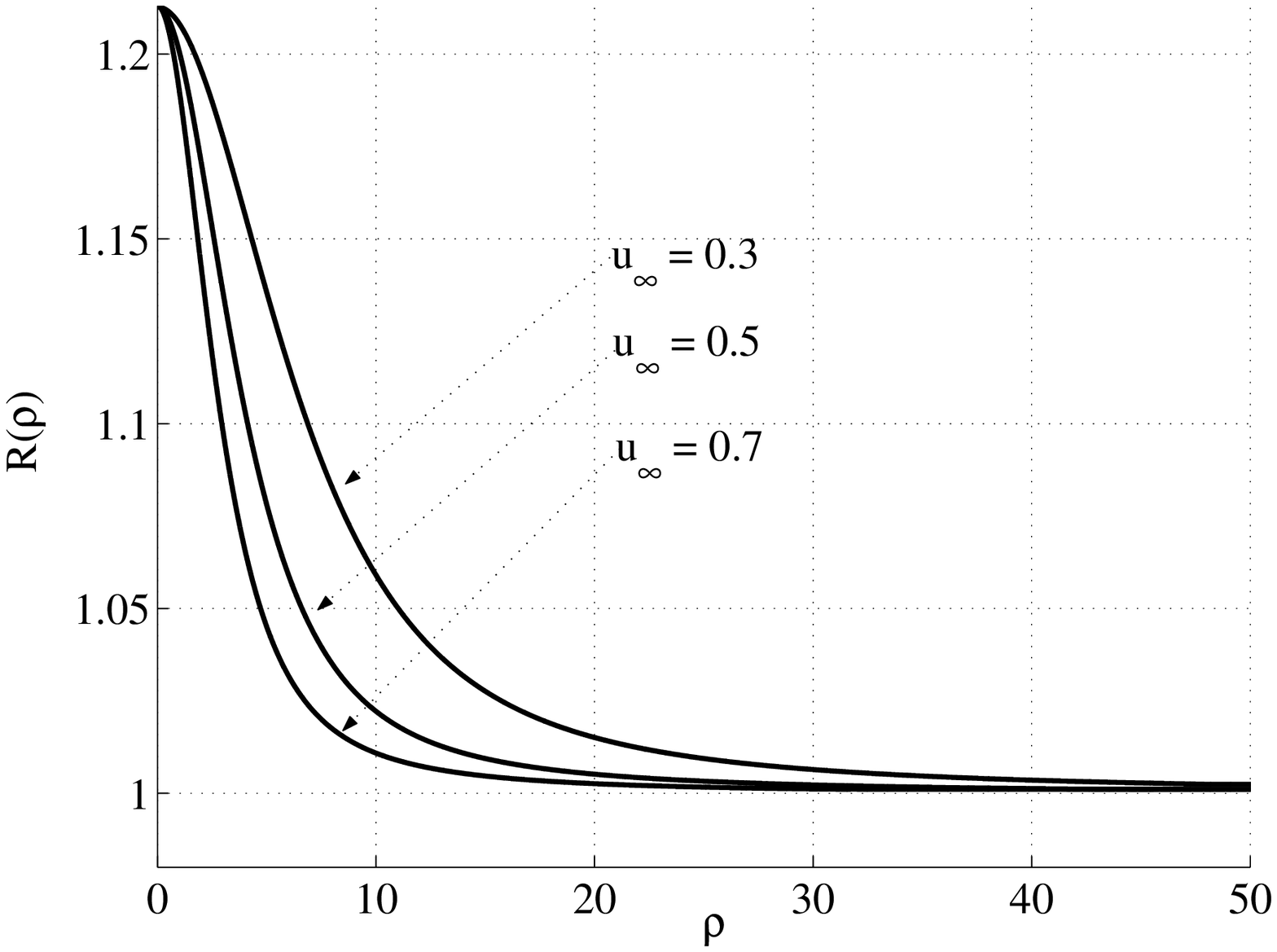}
\caption{}
\end{figure}
\end{document}